\documentclass[a4paper,fleqn,usenatbib]{mnras}

\hypersetup{draft}
\usepackage[T1]{fontenc}
\usepackage{ae,aecompl}
\usepackage{graphicx}
\usepackage{amsmath}
\usepackage{amssymb}	
\usepackage{euclid}
\usepackage{url}

\title[\Euclid colour selection]{\Euclid: The selection of quiescent and star-forming galaxies using observed colours\thanks{This paper is published on behalf of the Euclid Consortium.}}
\date{}

\author[L.Bisigello et al.]{L.~Bisigello$^{1,2}$\thanks{laura.bisigello@inaf.it}, U.~Kuchner$^{1}$, C.J.~Conselice$^{1}$, S.~Andreon$^{3}$, M.~Bolzonella$^{2}$, P.-A.~Duc$^{4}$,\newauthor
B.~Garilli$^{5}$, A.~Humphrey$^{6}$, C.~Maraston$^{7}$, M.~Moresco$^{2,8}$, L.~Pozzetti$^{2}$, C.~Tortora$^{9}$,\newauthor
G.~Zamorani$^{2}$, N.~Auricchio$^{2}$, J.~Brinchmann$^{6}$, V.~Capobianco$^{10}$, J.~Carretero$^{11}$,\newauthor
F.J.~Castander$^{12,13}$, M.~Castellano$^{14}$, S.~Cavuoti$^{15,16,17}$, A.~Cimatti$^{8,9}$, R.~Cledassou$^{18}$,\newauthor
G.~Congedo$^{19}$, L.~Conversi$^{20}$, L.~Corcione$^{10}$, M.S.~Cropper$^{21}$, S.~Dusini$^{22}$, M.~Frailis$^{23}$,\newauthor
E.~Franceschi$^{2}$, P.~Franzetti$^{5}$, M.~Fumana$^{5}$, F.~Hormuth$^{24}$, H.~Israel$^{25}$, K.~Jahnke$^{26}$,\newauthor
S.~Kermiche$^{27}$, T.~Kitching$^{21}$, R.~Kohley$^{20}$, B.~Kubik$^{28}$, M.~Kunz$^{29}$, O.~Le F\`evre$^{30}$,\newauthor
S.~Ligori$^{10}$, P.B.~Lilje$^{31}$, I.~Lloro$^{12,13}$, E.~Maiorano$^{2}$, O.~Marggraf$^{32}$, R.~Massey$^{33}$,\newauthor
D.C.~Masters$^{34}$, S.~Mei$^{35,36}$, Y.~Mellier$^{37,38}$, G.~Meylan$^{39}$, C.~Padilla$^{11}$, S.~Paltani$^{40}$,\newauthor
F.~Pasian$^{23}$, V.~Pettorino$^{41}$, S.~Pires$^{41}$, G.~Polenta$^{42}$, M.~Poncet$^{18}$, F.~Raison$^{43}$,\newauthor
J.~Rhodes$^{34}$, M.~Roncarelli$^{2,8}$, E.~Rossetti$^{8}$, R.~Saglia$^{25,43}$, M.~Sauvage$^{41}$,
P.~Schneider$^{32}$,\newauthor
A.~Secroun$^{27}$, S.~Serrano$^{12,44}$, F.~Sureau$^{41}$, A.N.~Taylor$^{19}$, I.~Tereno$^{45,46}$,\newauthor
R.~Toledo-Moreo$^{47}$, L.~Valenziano$^{2,48}$, Y.~Wang$^{49}$, M.~Wetzstein$^{43}$, J.~Zoubian$^{27}$
\\
    (Affiliations can be found after the references)
}

\date{Accepted 2020 March 12. Received 2020 March 10; in original form 2020 January 7}

\pubyear{2019}

\begin{document}
\label{firstpage}
\pagerange{\pageref{firstpage}--\pageref{lastpage}}
\maketitle
	\begin{abstract}
		The \Euclid mission will observe well over a billion galaxies out to $z\sim6$ and beyond. This will offer an unrivalled opportunity to investigate several key questions for understanding galaxy formation and evolution. The first step for many of these studies will be the selection of a sample of quiescent and star-forming galaxies, as is often done in the literature by using well known colour techniques such as the `$UVJ$' diagram. However, given the limited number of filters available for the \Euclid telescope, the recovery of such rest-frame colours will be challenging.
		We therefore investigate the use of observed \Euclid colours, on their own and together with ground-based \textit{u}-band observations, for selecting quiescent and star-forming galaxies. The most efficient colour combination, among the ones tested in this work, consists of the $(u-VIS)$ and $(VIS-J)$ colours. We find that this combination allows users to select a sample of quiescent galaxies complete to above $\sim70\%$ and with less than 15$\%$ contamination at redshifts in the range $0.75<z<1$. For galaxies at high-$z$ or without the $u$-band complementary observations, the $(VIS-Y)$ and $(J-H)$ colours represent a valid alternative, with $>65\%$ completeness level and contamination below 20$\%$ at $1<z<2$ for finding quiescent galaxies. In comparison, the sample of quiescent galaxies selected with the traditional $UVJ$ technique is only $\sim20\%$ complete at $z<3$, when recovering the rest-frame colours using mock \Euclid observations.  This shows that our new methodology is the most suitable one when only \Euclid bands, along with $u$-band imaging, are available.
		
	\end{abstract}
	
\section{Introduction}
	Galaxies show a clear bimodality in the distribution of their rest-frame ultraviolet and optical colours. Therefore, such colours are often considered when distinguishing and studying different galaxy populations \citep{Strateva2001,Blanton2003a,Baldry2004,Bell2004,Peng2010,Moresco2013,Jin2014,Fritz2014}. Because the optical spectrum of galaxies is dominated by the integrated light of their stellar population, any relation between their colours and magnitudes reflects differences in their star-formation histories, dust content, and metallicities. 

	In order to separate quiescent from star-forming galaxies  -- and thus galaxies with different star-formation histories -- with a simple but effective method, rest-frame $U-V$ colours have been extensively compared to the overall visible magnitude  \citep{Giallongo2005,Cassata2007,Labbe2007,Wyder2007,Jin2014,Lin2019}. However, galaxy observations at higher redshifts, e.g., $z\sim3$, require the addition of near-IR colours, that use, for example, the rest-frame  $J$ band, in order to distinguish between highly dusty, star-forming systems and quiescent galaxies \citep{Pozzetti2000, Wuyts2007}. As a consequence, the use of colour-colour diagrams such as the $UVJ$ technique has become a standard way to characterise galaxy populations and to study how they evolve through time \citep[e.g.,][]{Mendel2015,Fang2018}.
	The rest-frame $(U-V)$ and $(V-J)$ colours of galaxies have furthermore been demonstrated to evolve minimally with redshift \citep{Williams2009,Whitaker2011}.   Although the rest-frame colours of galaxies are highly dependent on the spectral energy distribution (SED) modelling, overall, they can be considered sufficiently accurate for normal galaxies if multiple bands are available.\par 
	
	\Euclid\footnote{\url{http://sci.esa.int/euclid/}} is a European Space Agency mission with the aim of mapping the geometry of the Universe and studying the evolution of cosmic structures and the distance-redshift relation. In order to achieve this goal, \Euclid will derive precise shapes and redshift measurement for over a billion galaxies out of $z\sim3$ and it will observe several millions galaxies out of $z\sim6$. \Euclid has a 1.2\,m primary mirror and two instruments on board. The  visible ($VIS$) instrument will provide high-quality visible imaging with an extremely wide broad-band filter covering between 550 and 900 nm and a mean image quality of $\sim0.{\rm\arcsec}23$ \citep{Cropper2010}. The complementary Near Infrared Spectrometer and Photometer (NISP) instrument will cover wavelengths from 900 to 2000\,nm with three broad-band filters, i.e., $Y$, $J$, and $H$  (see \autoref{fig:filters}), and a low-resolution slitless spectrometer \citep{Schweitzer2010}. The \Euclid Wide Survey is expected to cover 15\,000 deg$^2$ down to $10\sigma$ depth of 24.5\,mag in the visible filter and down to a $5\sigma$ depth of 24.0\,mag at near-infrared wavelengths. A deep survey two magnitudes deeper than the main survey will also be conducted over 40 deg$^2$ in the \Euclid Deep Fields.
	In addition to these main \Euclid surveys, extensive plans are in place to complement \Euclid observations with ground-based data from the ultraviolet to visible light \citep{Laureijs2010,Ibata2017} in order to improve the sampling quality of the SED for each galaxy. This is of course very challenging, given that the goal is to observe uniformly almost the entire extra-galactic sky at \Euclid depth, using ground-based instruments. 
	
	Overall, this extraordinary galaxy survey will be crucial not only for cosmological studies, but also to investigate several Legacy science key questions, especially related to galaxy formation and evolution.  Given that quiescent and star-forming galaxies represent the two most common evolutionary phases of galaxies, and considering the large amount of galaxies that will be observed by \Euclid,  it is essential to obtain a fast and reliable criterion to select quiescent and star-forming galaxies with the \Euclid photometric capability, as this will be the first step for many future studies. One of the dominant difficulties for this endeavour is the main \Euclid filter, $VIS$: its uncommonly large wavelength range was especially designed for \Euclid and has therefore never been used or tested with real data (see \autoref{tab:magdepth}). It is important to fully characterise the use of this filter for galaxy evolution studies, and a central part of this is testing its ability to distinguish between star-forming and passive galaxies. \par
	
	The aim of this work is therefore to utilise a set of mock \Euclid observations to analyse the efficiency of different \Euclid observed colours for separating quiescent and star-forming galaxies. The structure of the paper is the following: in \autoref{sec:data} we describe the derivation of the mock observations following three different methods. In \autoref{sec:qselection} we report the quiescent galaxies selection and the use of the standard rest-frame $U$, $V$, and $J$ colours to separate star-forming and quiescent galaxies. The capability of the different \Euclid observed colour combinations on isolating quiescent galaxies is then evaluated in \autoref{sec:colours}. We summarise our main finding in \autoref{sec:conclusions}.
	
	Throughout this paper, we use a Chabrier initial mass function \citep{Chabrier2003}, and a $\Lambda$CDM cosmology with $H_0=70\,\kmsMpc $, $\Omega_{\rm m}=0.27$, $\Omega_\Lambda=0.73$ and all magnitudes are in the AB system \citep{Oke1983}.
	
	
\section{Mock observations}\label{sec:data}
	\begin{table}
		\caption{$10\sigma$ depth in AB magnitude, central wavelength and full width half maximum (FWHM) of the four \Euclid filters and the CFIS/$u$ bands. The Deep Survey will be two magnitudes deeper than the primary survey in all bands.} 
		\centering 
		\begin{tabular}{c c c c}
			\hline\hline 
			\\
			band & $10\sigma$ depth & central wavelength [\AA] & FWHM [\AA]\\
			\hline
			$VIS$	&  24.50 & 7150 & 3640\\
			NISP/$Y$	&  23.24 & 10850 & 2660\\
			NISP/$J$	&  23.24 & 13750 & 4040\\
			NISP/$H$	&  23.24 & 17725 & 5020\\
			CFSI/$u$  & 24.20 & 3715 & 510\\
			\hline
		\end{tabular}
		\label{tab:magdepth}
	\end{table}
	\begin{figure}
		\centering
		\includegraphics[width=1\linewidth, keepaspectratio]{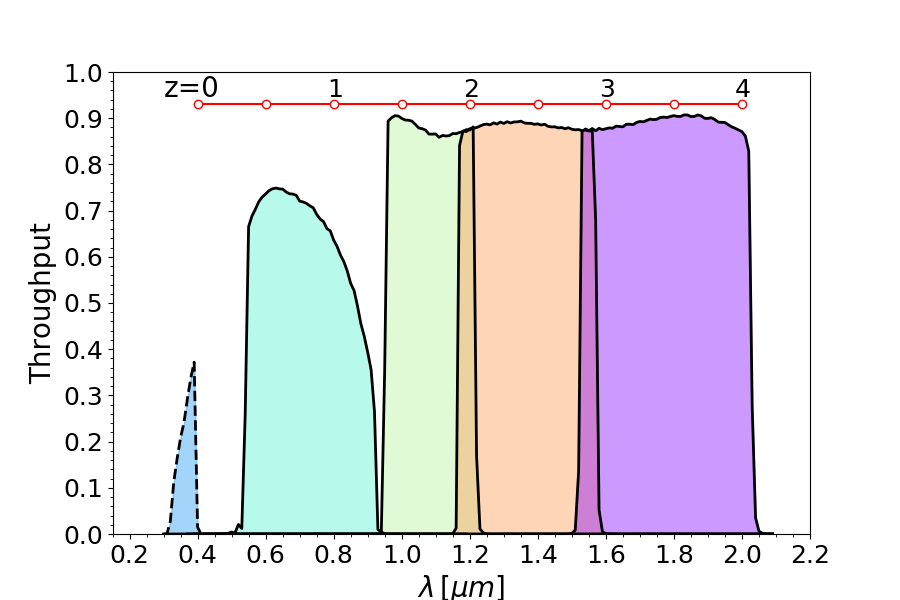}
		\caption{Throughput of the four main \Euclid filters (\textit{coloured regions and solid black lines}). From left to right, these are the $VIS$ filter, and the NISP/$Y$, NISP/$J$, and NISP/$H$ filters. We also include the throughput of the CFSI/$u$ band filter (\textit{blue region, dashed black line}). The red dots indicate the observed wavelength of the 4000\AA-break at different redshifts.}
		\label{fig:filters}
	\end{figure}
	
	\begin{table*}
		\caption{Summary of the different types of simulated data used in this work.} \label{tab:data sets}
		\begin{tabular}{cccc}
			\hline\hline 
			Name & Origin & N$_{\rm objects}$ & N$_{\rm quiescent}$\\
			\hline
			SED Wide & SED fitting from COSMOS2015 & 3\,249\,101 & 213\,837  \\
			SED Deep & SED fitting from COSMOS2015 & 5\,121\,526 & 303\,761  \\
			Int Wide & Interpolation from COSMOS2015 & 315\,755 & 21\,988 \\
			Int Deep & Interpolation from COSMOS2015 & 517\,890 & 30\,990  \\
			Flag Wide & Euclid Flagship mock galaxy catalogue & 12\,982 & 2\,576 \\
			Flag Deep & Euclid Flagship mock galaxy catalogue & 45\,162 & 3\,050\\
			\hline
		\end{tabular}
	\end{table*}

	We derive mock observations for the four broad-band filters on board \Euclid which are the visible $VIS$ filter and the NISP instrument's $Y$, $J$, $H$ filters. To test colour selections with a greater wavelength coverage, we also include the \textit{u}-band from the Canada-France Imaging Survey (CFIS) in our analysis. This band, as well as other ground-based optical bands such as the similar \textit{u}-band from the Large Synoptic Survey Telescope \citep[LSST;][]{Ivezic2008}, will be available over a large fraction of the fields (around 2/3 of \Euclid sky for CFIS) in order to complement \Euclid observations \citep{Ibata2017}. The 5 filter throughputs we consider are shown in \autoref{fig:filters} and the central wavelengths and widths are reported in \autoref{tab:magdepth}. Additional improvements can be expected if all 5 ancillary broad-bands ($u$, $g$, $r$, $i$ and,$z$) are available. However our present work focuses on the capability of the \Euclid mission alone. While ancillary data will become available, it will not be homogeneous and may not cover the full area observed by \Euclid.\par

	We derive fluxes for real and simulated galaxies in these bands using three different approaches that are summarised in \autoref{tab:data sets}. Two of these methods are based on real galaxies observed with current facilities and taken from the Cosmos Evolution Survey \citep[COSMOS,][]{Scoville2007}, while the third one is based on the Euclid Flagship mock galaxy catalogue based on theoretical SEDs. In all cases we consider separately the observational depth expected for the \Euclid Wide Survey as well as the \Euclid Deep Survey, which will reach two magnitudes deeper (see \autoref{tab:magdepth}). The magnitude distributions of all three data sets are compared in Appendix \ref{sec:mag_data sets}.
	
	
\subsection{Mock \Euclid fluxes from real galaxies}\label{sec:colours_rg}
	
	We start our work from the public  COSMSOS2015 catalogue \citep{Laigle2016} which contains multi-wavelength observations of more than a million objects over the 2 deg$^{2}$ of the COSMOS field. From the COSMOS2015 catalogue we consider 30 bands, reaching from the GALEX \citep{Zamojski2007} near-UV filter around 0.23$\,\micron$ to the \textit{Spitzer}/IRAC band at 4.5$\,\micron$ \citep{Sanders2007}.  We use aperture magnitudes measured within 3 arcsec and correct for photometric offsets, systematic offsets and galactic extinction, as suggested in \citet{Laigle2016}. Briefly, the first offset is derived from photometric data to correct for the incompleteness in the flux measured inside the fixed aperture. The second one is obtained by comparing the observed colours with the colours predicted with several theoretical templates, i.e. templates from \citet{Polletta2007} and \citet{Bruzual2003}, for a sample of galaxies with spectroscopic redshifts. The galactic extinction includes the foreground extinction derived by \citet{Allen1976}. We remove from the sample objects that are flagged as having inadequate optical photometry (\texttt{FLAGPETER}$>0$) and objects that are labelled as stars or X-ray sources in the COSMOS2015 catalogue. The 3673 X-ray sources in the catalogue are mainly active galactic nuclei but account for only a small fraction of sources compared to the final galaxy population. However, a similar selection should always be considered before applying the criteria we offer in this paper to future \Euclid samples. The final catalogue consists of 518\,404 galaxies  with photometric redshifts up to z$\sim$6.  \par
	
	For all the galaxies in the catalogue, we derive mock fluxes and magnitudes for the $VIS$, $Y$, $J$, and $H$ \Euclid bands and the CFIS/$u$ filter using two different approaches and considering the observational depth expected both for the \Euclid Wide and \Euclid Deep Surveys. However, the COSMOS2015 catalog is significantly shallower than the \Euclid Deep Survey, therefore many faint galaxies that will be detected in the \Euclid Deep Survey are missing in this catalogue.  \par
	
	\subsubsection{The \textit{Int} data set}
	The first method  to derive \Euclid mock observations is based on a linear interpolation of the 30 broad-band filters available in the COSMOS2015 catalogue. In particular, we use a broken line that connects the available COSMOS2015 observations as a proxy of each galaxy spectrum. We then interpolate this broken
	``spectrum" with the \Euclid filter throughputs. For the $J$, $Y$ and, $H$ filters this method is similar to interpolating the adjacent observed filters, but the described method is necessary to achieve a correct estimate for observations in the wide $VIS$ band. We do not include additional scatter to mimic the expected \Euclid photometric errors, because  the observational depth of the COSMOS2015 catalogue is similar or shallower than the one expected for the \Euclid Surveys. For example, the observed magnitude errors in the COSMOS2015 $J$ ($Y$) band are, on average, 1.5 (3) times larger than the magnitude errors expected for the \Euclid $J$ ($Y$) filter, assuming the observational depth of the \Euclid Wide Survey. On the other hand, magnitude errors in the COSMOS2015 $H$ band are similar to the expected \Euclid $H$ band errors for the \Euclid Wide Survey, showing that the two surveys are comparable in this band.
	Hereafter we refer to mock observations derived by using this method based on the 518\,404 galaxies selected from the COSMOS2015 catalogue as data sets \textit{Int Wide} and \textit{Int Deep}, depending on the assumed observational depth. Finally, we select only galaxies with S/N$>$3 in the $VIS$ band, which leads to 315\,755 galaxies in our \textit{Int Wide} sample and 517\,890 galaxies in our \textit{Int Deep} sample. \par
	
	\subsubsection{The \textit{SED} data set}
	For the second approach, we derive mock observations from the best theoretical template that describes the SED of each galaxy. For this, we use the observations in 30 filters of the COSMOS2015 catalogue. In particular, we use the public code \texttt{LePhare} \citep{Arnouts1999,Ilbert2006} and consider \citet{Bruzual2003} templates with solar and subsolar (0.008 Z$_{\odot}$) metallicities, exponentially declining star formation histories with timescale $\tau$ between 0.1 and 10 Gyr, ages between 0.1 and 12 Gyr, the \citet{Calzetti2000} reddening law, and 12 values of colour excess between 0 and 1. We did not apply any cut in S/N on the observed COSMOS2015 observations and we considered magnitude errors and upper limits as derived by \citet{Laigle2016}. We only apply a lower limit to the magnitude errors, i.e. 0.01 mag, in order to avoid the fit being driven by single observations. We only consider  exponentially declining star-formation histories, since they generally describe the bulk of the quiescent galaxy population at z$<$3 well. We will get back to this later, when we compare results of the SED, \textit{Int Wide} and \textit{Int Deep} data sets, where we used different assumptions concerning the star-formation history.   \par 
	
	We also allow the code to add nebular emission lines, as explained in \citet{Ilbert2006}. Note that the effect of including nebular emission lines in the fit is minor, given that this work focuses on galaxies at z<3 and nebular emission lines are more prominent in high-z galaxies \citep{Fumagalli2012,Duncan2014,Marmol-Queralto2016}. Moreover, equivalent widths higher than $\sim350\,$\AA,\mbox{$\, \sim260\,$\AA,}$\,\sim390\,$\AA$\,$ and, $\sim480\,$\AA$\,$ are necessary to produce a detectable boost ($\Delta Y>$0.1 mag) in the $VIS$, $Y$, $J$ and, $H$ filters, respectively. In addition, during the fit we fix the redshift to the value reported in the COSMOS2015 catalogue and the age of each galaxy is constrained to be smaller than the age of the Universe at the galaxy's redshift. \par
	After deriving the best SED templates, we randomise each flux 10 times using a normal distribution centred on the flux value and with a standard deviation equal to the expected flux error. This depends on the assumed survey depth and is defined as one tenth of the flux corresponding to a S/N$=$10. Note that this is equal to 24.50 (26.50) AB\,mag in the $VIS$ band for the Wide (Deep) Survey (see \autoref{tab:magdepth} for the depth in each filter).  Hereafter we refer to mock observations derived using this method as data set \textit{SED Wide} or \textit{SED Deep}, depending on the assumed observational depth. We remove from the final catalogues every galaxy which has S/N$<$3 in the $VIS$ filter. The data set \textit{SED Wide} consists of 3\,249\,101 mock galaxies, while the \textit{SED Deep} catalogue contains 5\,121\,526 mock galaxies.\par

	We also infer rest-frame $U$, $V$, and $J$ magnitudes and the specific star formation rate (sSFR) of each galaxy from the best SED template. To derive rest-frame $U$, $V$, and $J$ magnitudes, we consider $U$ and $V$ band-passes from \citet{MaizApellaniz2006}  and the $J$ band-pass from the Two Micron All Sky Survey \citep{Skrutskie2006}. $U$, $V$, and $J$ rest-frame magnitudes derived in this work are consistent with those reported in the COSMOS2015 catalogue. Note that we chose to re-calculate these rest-frame colours for consistency, since we later in the paper present the same rest-frame colours derived using the \Euclid mock observations. sSFR derived in this way are considered as the \textit{true} sSFR associated with each galaxy in the \textit{SED} and \textit{Int} data sets. Moreover, for the rest of the paper, we assign to each galaxy its \textit{true} redshift. This corresponds to the redshift of the SED template derived from the real observations (used to infer the \Euclid mock observations in our work). However, we assume it will be possible to recover photometric redshifts with an accuracy good enough for the redshift bins considered here, i.e., $\sigma_{z}=$0.25 or 0.5 at z$>$1.5. This is more than realistic given that the requirement to perform \Euclid cosmological studies is to obtain a photometric redshift accuracy of $\sigma_{z}<0.05\,(1+z)$.
	\par
	The two methods described in this section are complementary. The first one depends on the observed COSMOS2015 photometric errors, which may not completely match the future \Euclid photometric uncertainties. It also uses a few model assumptions (i.e., the photometric offsets are derived from theoretical templates). The second method depends on the theoretical templates, reddening law, and star formation histories used for the SED fit, but matches the expected \Euclid photometric errors. The data sets differ in galaxy numbers because of the adopted \Euclid Survey depth and the different approaches used for including photometric errors. We remind the reader that we randomise 10 times the observed galaxies in the \textit{SED} data sets to mimic the expected \Euclid photometric errors. On the other hand, we did not randomise the fluxes in the \textit{Int} data sets because the COSMOS2015 photometric errors already influence the broken ``spectrum" used to derive the mock observations.  \par

\subsection{Mock \Euclid fluxes from simulations}\label{sec:2}

	We complete our data sets with mock observations obtained from 
the Euclid internal Scientific Challenge (SC456) that make use of galaxy properties based on the Euclid Flagship mock galaxy catalogue v1.7.17. This mock 
catalogue populates the Flagship dark matter simulation \citep{Potter2017} 
with galaxies following similar recipes to those implemented in the 
MICE mock catalogues\footnote{\url{http://www.ice.csic.es/en/content/68/mice-simulations}} \citep{Carretero2015,Fosalba2015a,Fosalba2015b,Crocce2015}. The Flagship simulation was designed to mimic the observational depth and conditions of the actual \Euclid survey (Castander et al., in prep).  It is therefore a theoretical determination which complements our observational inference of colours described in the previous section. Adding simulated galaxies with known input parameters to our analysis offers the advantage of providing full control over measurement errors while minimising systematic errors.\par
The simulation catalogue was generated using a hybrid Halo Occupation Distribution 
and Halo Abundance Matching prescriptions to populate the Flagship Friends of Friends dark matter halos. The Flagship simulation used the following cosmological parameters: $\Omega _{\rm m}=0.319$, $\sigma_{8} = 0.83$, $n_{\rm s}=0.96$, $\Omega _{\rm b}=0.049$, $\Omega_{ \Lambda}=0.681$, and $h=0.67$. These values of $\Omega _{\rm m}$ and $\Omega_{ \Lambda}$ are slightly different from those used in the creation of the other mock observations, but the impact is negligible on our results as they do not influence galaxy colours. \par
The catalogue was built to follow a number of local observational constraints, among which are i) the luminosity function at $z=0.1$ \citep{Blanton2003}, ii) the galaxy clustering as a function of luminosity and colour as observed in the Sloan Digital Sky Survey up to $z=0.25$ \citep{Zehavi2011} and iii) the colour-magnitude diagram of galaxies at $z<0.3$ \citep{Blanton2005}. A template taken from the SED library of \citet{Ilbert2009} is associated to each galaxy in the simulation. This library includes templates from \citet{Bruzual2003}, with ages ranging from 3 to 0.03 Gyr, and template for elliptical and spiral galaxies are taken from \citet{Polletta2007}.
The final Euclid Flagship mock galaxy catalogue v1.7.17 contains galaxies up to redshifts $z=2.3$ with \Euclid H-band apparent magnitudes down to $H\sim26$ mag.
    
We include photometric errors for these galaxies by randomising each flux 10 times by considering a normal error distribution centred on the real value with a standard deviation equal to the noise expected for the \Euclid Wide Survey and the \Euclid Deep Survey, respectively (see Table \ref{tab:magdepth}). The Euclid Flagship mock galaxy catalogue has a restricted number of quiescent galaxies with detections in the $u$ band, therefore this data set is not used to derive colour selections which include the $u$ band. 
Hereafter we refer to mock observations derived by using this method as data set \textit{Flag Wide} and \textit{Flag Deep}, depending on the assumed observational depth. Both data sets are created from a sample of 80\,790 mock galaxies limited to z$<$2.3. Because of the completeness of the  Euclid Flagship mock galaxy catalogue, the mock catalogue \textit{Flag Deep} created in this work is missing part of the population of faint galaxies expected in the \Euclid Deep Survey. \par
	
A general comparison of the properties of the $Flag$, $Int$, and $SED$ Wide data sets is presented in Appendix \ref{sec:mag_data sets}.

\section{Quiescent galaxies initial selection}\label{sec:qselection}
	\begin{figure}
		\centering
		\includegraphics[width=0.98\linewidth, keepaspectratio]{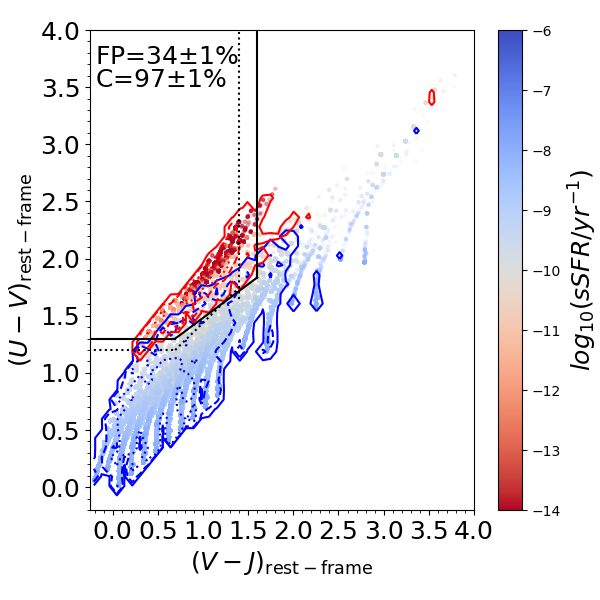}
		\caption{$(U-V)$ and $(V-J)$ rest-frame colours derived from the best SED template describing 518\,404 galaxies with 30 COSMOS2015 bands. Boundaries that select quiescent galaxies are taken from \citet{Whitaker2011} and are shown for $z=0$ as \textit{black solid lines} and $z=3$ as \textit{black dotted lines}. Galaxies are colour-coded depending on their sSFR. The blue and red contours show 99.7$\%$ (\textit{solid lines}), 95$\%$ (\textit{dashed lines}) and 68$\%$ (\textit{dotted lines}) of the number density of star-forming [$\logten({\rm sSFR/\,yr^{-1}})>-10.5$] and quiescent galaxies [$\logten({\rm sSFR/\,yr^{-1}})<-10.5$], respectively. On the top left, we report the completeness (C) and false-positive fraction (FP) of the quiescent galaxy selection with the corresponding Poisson errors.}
		\label{fig:UVJ_original}
	\end{figure}	
	
	In this section, we first describe our initial selection of quiescent and star-forming galaxies with a rest-frame $UVJ$ selection. Then, we compare this reference selection with selections that use \Euclid filters only: once to derive $U$, $V$, and $J$ rest-frame colours and once to derive sSFRs. \par
	In the literature, several studies have identified quiescent galaxies using a fixed threshold in sSFR. However, this threshold is not uniform and varies depending on the properties of the data set and how the star formation rate and masses are measured. \citep{McGee2011,Wetzel2012,Lin2014}, e.g., on the minimum  of the bimodal distribution of the sSFRs of galaxies at low redshift \citep{Kauffmann2004,Wetzel2013,Renzini2015,Bisigello2018}. \par
	
	In the following, we define star-forming galaxies as objects with 
	\begin{equation*}
	\logten({\rm sSFR/\,yr^{-1}})>-10.5\,,
	\end{equation*} 
	while quiescent galaxies have 
	\begin{equation*}
	\logten({\rm sSFR/\,yr^{-1}})<-10.5\,.
	\label{eq:eq_sSFR}
	\end{equation*} 
    \noindent For the initial selection in the data sets \textit{SED} and \textit{Int}, we obtain the sSFR of each galaxy from the SED template that best describes and fits the 30 bands of the COSMOS2015 catalogue.
	As mentioned before, mock observations derived from the Euclid Flagship mock galaxy catalogue (data sets \textit{Flag}) do not include a sufficient number of galaxies with detection in the CFIS/$u$-band filter and, therefore, for these data sets we limit our analysis to colours of the $VIS$ and NISP filters. The sSFR for these data sets is taken from the Euclid Flagship mock galaxy catalogue.\par
	Throughout the paper, we test the different selection criteria by comparing them with the above-mentioned selection of quiescent galaxies from the observations in the 30 COSMOS2015 bands or the Euclid Flagship mock galaxy catalogue. The number of quiescent galaxies in each data set is reported in \autoref{tab:data sets}. We evaluate the different methods to derive quiescent galaxies considering three different quantities. 
	\begin{enumerate}
	    \item The mixing of quiescent and star-forming galaxies. This is defined as the percentage of galaxies inside the intersection between the areas containing 68$\%$ of both populations, looking at their number density distributions in colour space.
	    \item The completeness (C). This consists of the fraction of quiescent (or star-forming) galaxies that is correctly recognised by the analysed selection criteria.
	    \item The false-positive fraction (FP). This is the fraction of star-forming galaxies that are wrongly identified as quiescent by the analysed selection criteria, or vice-versa, the fraction of quiescent galaxies that is erroneously identified as star-forming. For readers more familiar with the concept of purity, this is equivalent to 1-FP.  
\end{enumerate}
\par
	
	As a first test, we compare the rest-frame colours $(U-V)$ and $(V-J)$ with the sSFR, both taken from the COSMOS2015 catalogue. We do this to verify our initial selection of quiescent galaxies. Since the $(U-V)$ and $(V-J)$ colour selection was derived from the empirical galaxy SED, we expect the two methods to be broadly consistent. Indeed, \autoref{fig:UVJ_original} shows that there is little mixing of star-forming and quiescent galaxies in the $UVJ$ plane and that they are well separated by the criteria described in \citet{Whitaker2011}: black solid lines for $z=0$ and dotted lines for $z=3$. Overall, the sSFR and $UVJ$ selections agree for 97$\%$ of quiescent galaxies. However, 34$\%$ of all star-forming galaxies are misclassified. Most of the misclassified galaxies have low star-formation rates, on average $\logten({\rm sSFR/\,yr^{-1}})\sim-$10.2, which means that that they are close to the boundary separating quiescent from star-forming galaxies. This test confirms that the majority of quiescent galaxies selected with the specified cut in sSFR is consistent with a selection using $UVJ$ colours.\par 

\subsection{Deriving $U$,$V$, and $J$ rest-frame colours and sSFR with \Euclid}\label{sec:UVJ_Euclid}

	\begin{figure}
		\centering
		\includegraphics[width=1\linewidth, keepaspectratio]{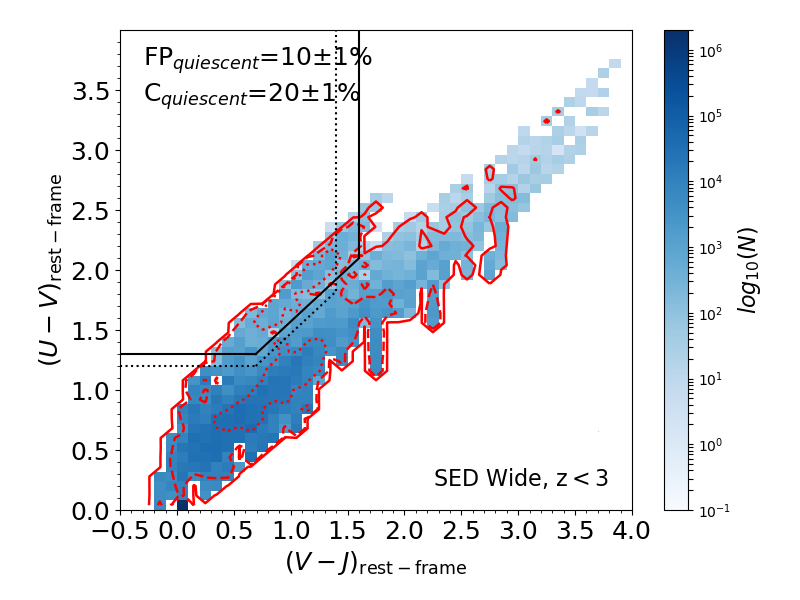}
		\caption{$(U-V)$ and $(V-J)$ rest-frame colours derived from the \Euclid filters $VIS$, $Y$, $J$ and $H$, considering the \textit{SED Wide} data set. As in \autoref{fig:UVJ_original}, the area containing quiescent galaxies is shown for $z=0$ in \textit{black solid lines} and $z=3$ in \textit{black dotted lines} \citep{Whitaker2011}. The red lines show the 99.7$\%$ (\textit{solid lines}), 95$\%$ (\textit{dashed lines}) and 68$\%$ (\textit{dotted lines}) contours of the number density of quiescent galaxies. For clarity, only the distribution of star-forming galaxies is shown in blue. This clearly shows the high contamination for quiescent galaxies. Star-forming and quiescent galaxies are selected using the rest-frame colours derived from the original 30 COSMOS2015 bands (\autoref{fig:UVJ_original}). On the top left, we report the false-positive fraction (FP) and the completeness (C) of the quiescent galaxy population with the corresponding Poisson errors.} 
		\label{fig:UVJ_Euclid}
	\end{figure}
	
	\begin{figure}
		\centering
		\includegraphics[width=1\linewidth, keepaspectratio]{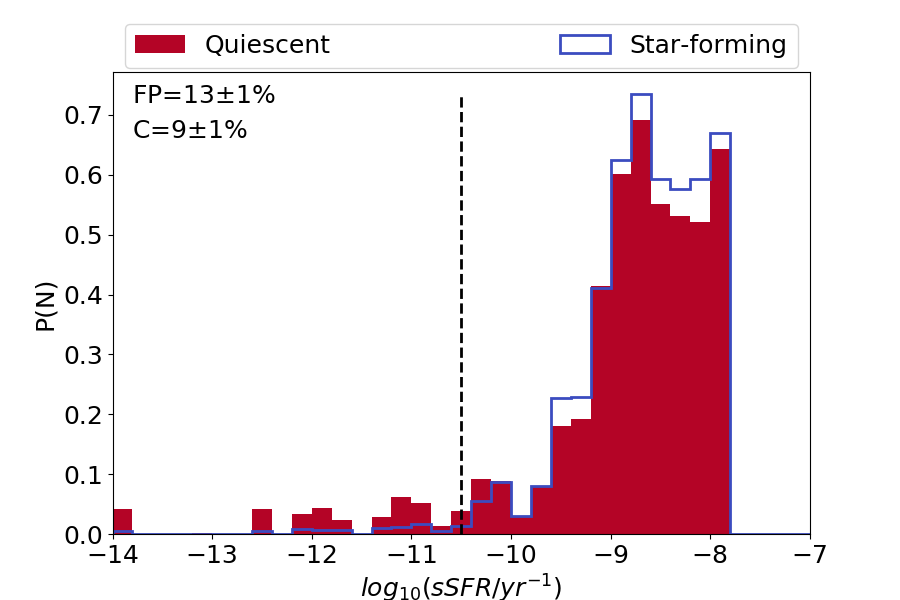}
		\caption{Distribution of the sSFR for galaxies in the \textit{SED Wide} data set, derived from the best SED template describing the four \Euclid band observations. The distribution is shown for galaxies that were classified as star-forming (\textit{empty blue histogram}) and quiescent galaxies (\textit{filled red histogram}) with the 30 COSMOS2015 filter observations -- our reference frame in this test. The dashed black vertical line shows the $\logten({\rm sSFR/\,yr^{-1}})=$-10.5 limit that we choose as the separation between quiescent and star-forming galaxies (see \autoref{sec:qselection}). The completeness (C) and false-positive fraction (FP) for the selection of quiescent galaxies is shown at the top left with the corresponding Poisson errors. Observations in only four filters are insufficient to recover the original SED with enough accuracy to properly predict the sSFR.}
		\label{fig:SED_sSFR}
	\end{figure}
	Following the success of the $UVJ$ colour combination to separate galaxy types in the original COSMOS2015 catalogue, we now investigate if it is possible to recover the correct rest-frame $(U-V)$ and $(V-J)$ colours from \Euclid observations.
	To derive the rest-frame colours with \Euclid observations, we apply the same method that we also used with the 30 COSMOS2015 bands (see \autoref{sec:colours_rg}): the algorithm searches for the theoretical SED template that best describes the four \Euclid mock observations. In this test, we allow the redshift to vary in the fit, similar to how future analyses with \Euclid observations will be done.  
	
	In \autoref{fig:UVJ_Euclid}, we show the $UVJ$ rest-frame selection derived from galaxies with the four \Euclid filters $VIS$, $Y$, $J$ and $H$, compared to our reference $UVJ$ rest-frame selection using the 30 COSMOS2015 bands. Reported results in this figure are for the \textit{SED Wide} data set.
	The majority of star-forming galaxies are correctly identified, as is evident from the high completeness (87$\%$) of the recovered star-forming population, and a relatively low false-positive fraction (10$\%$) of the quiescent galaxy population. 
	However, a very large fraction -- around 80$\%$ -- of quiescent galaxies are wrongly identified as star-forming galaxies. The results do not change much if we limit our analysis to z$<$1, as the completeness and false-positive fraction of quiescent galaxies are still 20$\%$ and 10$\%$, respectively. \par
	To better understand why we recover such low fractions of quiescent galaxies, we repeat the SED fit twice, each time slightly altering our approach. First, we fix the redshift to the ``true" redshift, rather than allowing the redshift to vary during the fitting process. In a second test, we adopt the photometric redshift precision expected for \Euclid, i.e. $\sigma_{z}<0.05\,(1+z)$. In the first case, both the completeness and false positive fraction for quiescent galaxies increase moderately from 20$\%$ and 10$\%$ to 41$\%$ and 31$\%$, respectively. We obtain similar results when we change the redshift errors to the photometric redshift precision of \Euclid, i.e., C$_{quiescent}=40\%$ and FP$_{quiescent}=32\%$. The moderate success of this test highlights the challenges that go along with recovering the correct SED template with only four \Euclid bands -- and therefore also for deriving the correct  $(U-V)$ and $(V-J)$ rest-frame colours -- even if high precision redshifts are available. 
	\medskip
	
	We further test whether it is possible to separate star-forming from quiescent galaxies with sSFR's derived from observations in the four \Euclid filters $VIS$, $Y$, $J$ and $H$. For this, we use the same SED templates that we used to derive the rest-frame colours to also retrieve 
	the sSFR's.
	In \autoref{fig:SED_sSFR} we show the recovered sSFR distribution for quiescent (red filled histogram) and star-forming (blue open histogram) galaxies of the \textit{SED Wide} data set. It is evident that observations in only four filters are insufficient to recover the original SED with enough accuracy to properly predict the sSFR's. In particular, almost all galaxies (both quiescent and star-forming) have sSFR's consistent with star-forming galaxies. Only 9$\%$ of the quiescent galaxy population is correctly identified, i.e., has $\logten({\rm sSFR/\,yr^{-1}})<-$10.5. At the same time, sSFR-selected quiescent galaxies contain 13$\%$ false-positives. It is difficult to recover the correct sSFR, but the redshift uncertainties cannot be solely responsible for this, since we have shown that the completeness of quiescent galaxies does not increase dramatically (only to 30$\%$), if we fix the redshift during the spectral fitting. We speculate that the choice of incorrect templates is likely responsible for the high incompleteness in recovering quiescent galaxies with accurate sSFRs.  \par
	
	In summary, we find that when only observations in the four \Euclid filters are available, neither the $(U-V)$ and $(V-J)$ rest-frame colours nor the sSFR are suitable to select quiescent galaxies with sufficient precision. In the rest of the paper we therefore test alternative methods to isolate quiescent galaxies with \Euclid observed colours.  
	
\section{Comparison of \Euclid colour combinations}\label{sec:colours}
	
	\begin{figure}
		\centering
		\includegraphics[width=1\linewidth, keepaspectratio]{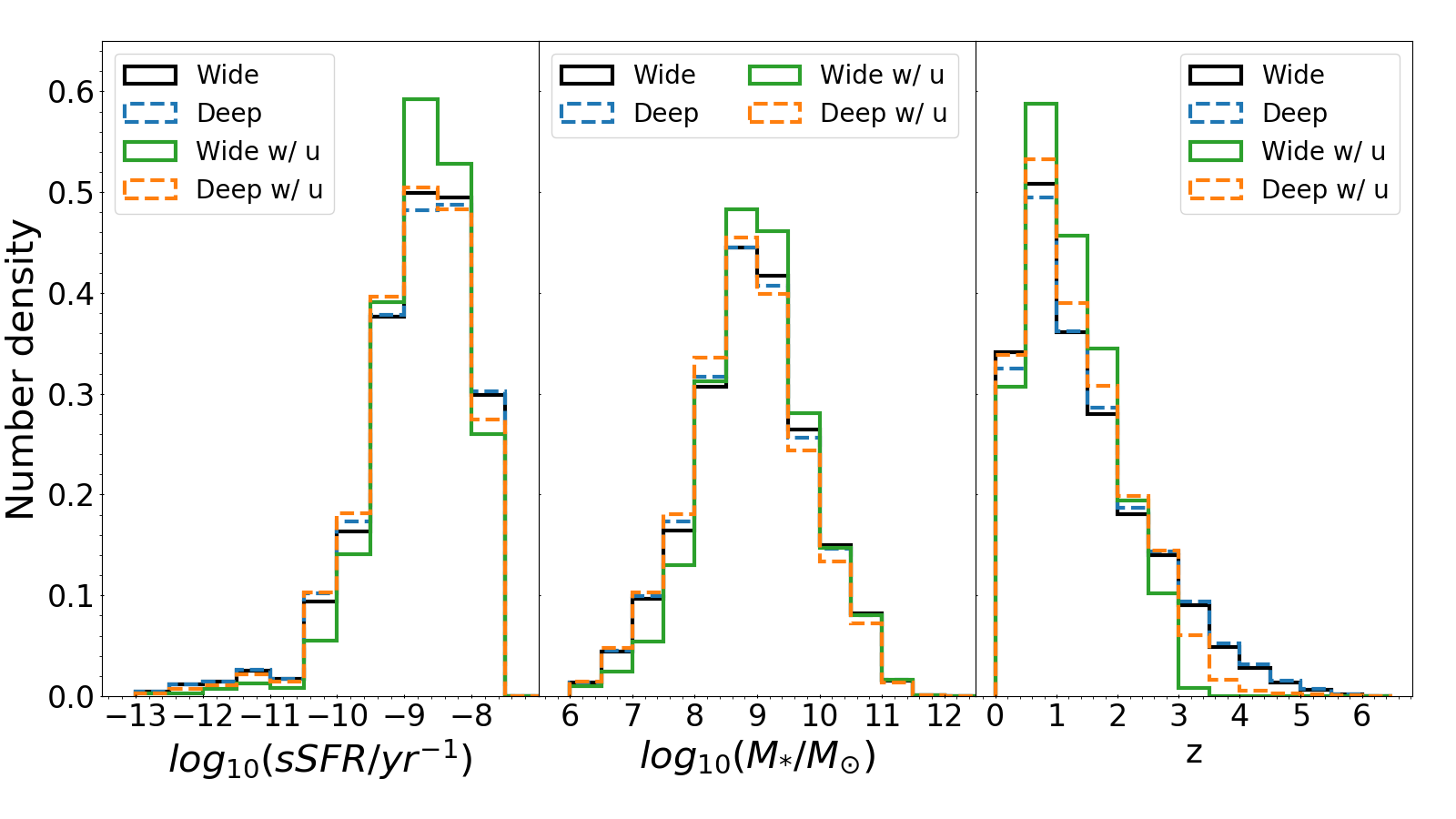}
		\caption{sSFR (\textit{left}), stellar mass (\textit{centre}) and redshift (\textit{right}) number density distribution of galaxies with $VIS$ observations in the \Euclid Wide (\textit{green solid lines}) and Deep Survey (\textit{orange dashed lines}), as well as for the subsample of galaxies with both $u$ and $VIS$ band observations in the \Euclid Wide (\textit{black solid lines}) and Deep Survey (\textit{blue dashed lines}). Results are shown for mock observations in the \textit{SED Wide} and \textit{Deep} data sets.}
		\label{fig:VIS_u_sample}
	\end{figure}

		\begin{table*}
		\caption{Fraction of star-forming and quiescent galaxies occupying the intersection between the areas containing 68$\%$ of the two populations in different colour space at z$<$3.} \label{tab:col_mixing}
		\begin{tabular}{ccccccccc}
			\hline\hline 
			Color & Population & $SED\,Wide$ & $SED\,Deep$ & $Int\,Wide$ & $Int\,Deep$ & $Flag\,Wide$ & $Flag\,Deep$ & Average\\ 
			\hline
			($VIS$ - $Y$) vs. ($Y$-$H$)  & quiescent    & 31$\%$ & 36$\%$ & 37$\%$ & 44$\%$ & 50$\%$ & 45$\%$ & 40.5$\%$\\
			                             & star-forming & 23$\%$ & 35$\%$ & 33$\%$ & 51$\%$ & 23$\%$ & 23$\%$ & 31.3$\%$\\
			 ($VIS$ - $Y$) vs. ($Y$-$J$) & quiescent    & 39$\%$ & 40$\%$ & 38$\%$ & 46$\%$ & 60$\%$ & 52$\%$ & 45.8$\%$\\
			                             & star-forming & 29$\%$ & 42$\%$ & 37$\%$ & 52$\%$ & 34$\%$ & 23$\%$ &36.2$\%$ \\
			 ($VIS$ - $J$) vs. ($J$-$H$) & quiescent    & 28$\%$ & 32$\%$ & 42$\%$ & 45$\%$ & 56$\%$ & 51$\%$ & 42.3$\%$ \\
			                             & star-forming & 20$\%$ & 36$\%$ & 33$\%$ & 52$\%$ & 27$\%$ & 22$\%$ &31.7$\%$ \\
			 ($VIS$ - $H$) vs. ($Y$-$J$) & quiescent    & 45$\%$ & 41$\%$ & 41$\%$ & 48$\%$ & 55$\%$ & 52$\%$ & 47.0$\%$\\
			                             & star-forming & 32$\%$ & 43$\%$ & 37$\%$ & 53$\%$ & 34$\%$ & 23$\%$ & 37.0$\%$\\
			 ($VIS$ - $Y$) vs. ($J$-$H$) & quiescent    & 30$\%$ & 31$\%$ & 25$\%$ & 44$\%$ & 55$\%$ & 47$\%$ & 38.7$\%$\\
			                             & star-forming & 19$\%$ & 30$\%$ & 30$\%$ & 50$\%$ & 32$\%$ & 26$\%$ & 31.2$\%$\\
			 ($u$ - $VIS$) vs. ($VIS$-$J$) & quiescent  & 0$\%$ & 0$\%$ & 0$\%$ & 15$\%$ & 0$\%$ & 0$\%$ & 2.5$\%$\\
			                               & star-forming & 0$\%$ & 0$\%$ & 0$\%$ & 40$\%$ & 0$\%$ & 0$\%$ & 6.7$\%$\\
			\hline
		\end{tabular}
	\end{table*}
		
	\begin{figure*}
		\centering
		\includegraphics[width=1\linewidth, keepaspectratio]{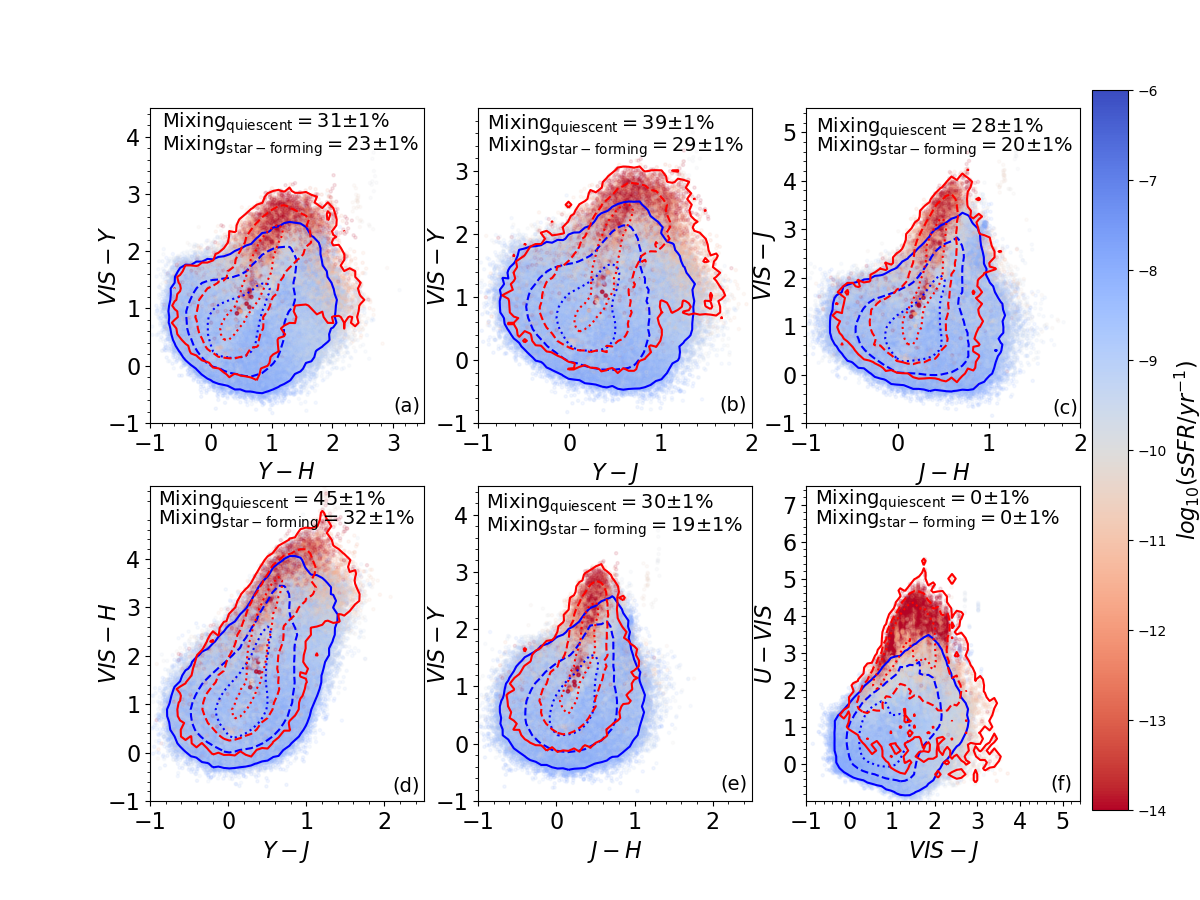}
		\caption{\Euclid observed colours for mock galaxies in the data set \textit{SED Wide} at z$<$3. The panels show different combinations of \Euclid observed colours. Galaxies are colour-coded depending on their original sSFR value (see text). The blue and red lines show the 99.7$\%$ (\textit{solid lines}), 95$\%$ (\textit{dashed lines}) and 68$\%$ (\textit{dotted lines}) contours of the number density of star-forming [$\logten({\rm sSFR/\,yr^{-1}})>-10.5$] and quiescent galaxies [$\logten({\rm sSFR/\,yr^{-1}})<-10.5$], respectively. In the top left of each panel we report the fraction of quiescent and star-forming galaxies occupying the intersection between the areas containing 68$\%$ of the two populations. The best separation between quiescent and star-forming galaxies is achieved with the $(u-VIS)$ and $(VIS-J)$ observed colour combination (lower right panel), which requires auxiliary data.}
		\label{fig:Obs_col}
	\end{figure*}

	We now investigate the ability to isolate quiescent galaxies from the star-forming galaxy population with various colour combinations available through \Euclid follow-up observations. For this we use \Euclid mock observations derived using the three methods described in the previous sections.  We limit our analysis to the use of aperture photometry, but the inclusion of morphological and spectroscopic information is expected to improve the purity of the sample \citep{Moresco2013, Andreon2018}. The addition of these features will be investigated in a future work. 
 To create a space that resembles the $UVJ$ plane, we first include the ground-based CFIS/$u$ band that will be available to complement \Euclid observation over much of the fields. Similar $u$-band filters will be available through LSST and other ground-based imaging surveys.
	
	In \autoref{fig:VIS_u_sample} we show the redshift, stellar mass and sSFR distributions of galaxies with $VIS$ observations (Wide and Deep) and the subsamples with both $u$-band and $VIS$ detections (Wide and Deep), considering the different observational depths expected for both filters in the two Surveys (see Table \ref{tab:magdepth}). Overall, around 63$\%$ (90$\%$) of galaxies in the \Euclid Wide (Deep) Survey with $VIS$ observations are detected in the $u$-band as well. Not surprisingly, the $u$-band observations limit the sample to low-redshift galaxies. In the \Euclid Wide Survey, they also  exclude some of the low-mass galaxies from the sample. In the future, it will be necessary to take into account this sample selection when considering colour criteria including the $u$-band filter.  \par 
	
	\autoref{fig:Obs_col} shows colour-colour plots of a variety of \Euclid colour combinations, including the $u$-band filter, for galaxies in the data set \textit{SED Wide}. The colours are derived from the best SED template obtained by including photometric errors, as explained in \autoref{sec:colours_rg}. Results are shown for mock galaxies up to z$=$3. Note that we found similar results in the other data sets, i.e., \textit{SED Deep}, \textit{Int Deep} and \textit{Int Wide} (\autoref{sec:data}), as listed in \autoref{tab:col_mixing}. 
	For each observed colour combination, we derive the percentage of quiescent and star-forming galaxies overlapping in colour-space, as this is an indication of the effectiveness of the method.  
	This is done by comparing the number density distribution of the quiescent and star-forming galaxy populations in each colour-space and then deriving the percentage of galaxies inside the intersection between the areas containing 68$\%$ of both populations. 
	
	The best separation between quiescent and star-forming galaxies is achieved with the $(u-VIS)$ and $(VIS-J)$ observed colour combination (\autoref{fig:Obs_col}, last panel). Using these colours, quiescent and star-forming galaxies overlap only outside the 68$\%$ areas. 
	In all other colour combinations a large fraction (more than 20$\%$) of quiescent and star-forming galaxies overlap in colour-space within the 68$\%$ areas. Among the \Euclid-only colour-combinations (i.e., that do not include the additional information of the $u$-band), the $(VIS-Y)$ vs. $(J-H)$ is most effective to separate populations. For this colour combination,  and considering the average among all data sets (see \autoref{tab:col_mixing}), the two galaxy populations have the smallest overlap -- even if only by a few percentage units.
	The real potential of the $(VIS-Y)$ vs. $(J-H)$ colour combination is revealed splitting the sample in redshift intervals, as will become obvious in the next sections.

	\subsubsection{Redshift separation: the $(u-VIS)$ and $(VIS-J)$ colours}
	
	\begin{figure*}
		\centering
		\includegraphics[width=1\linewidth, keepaspectratio]{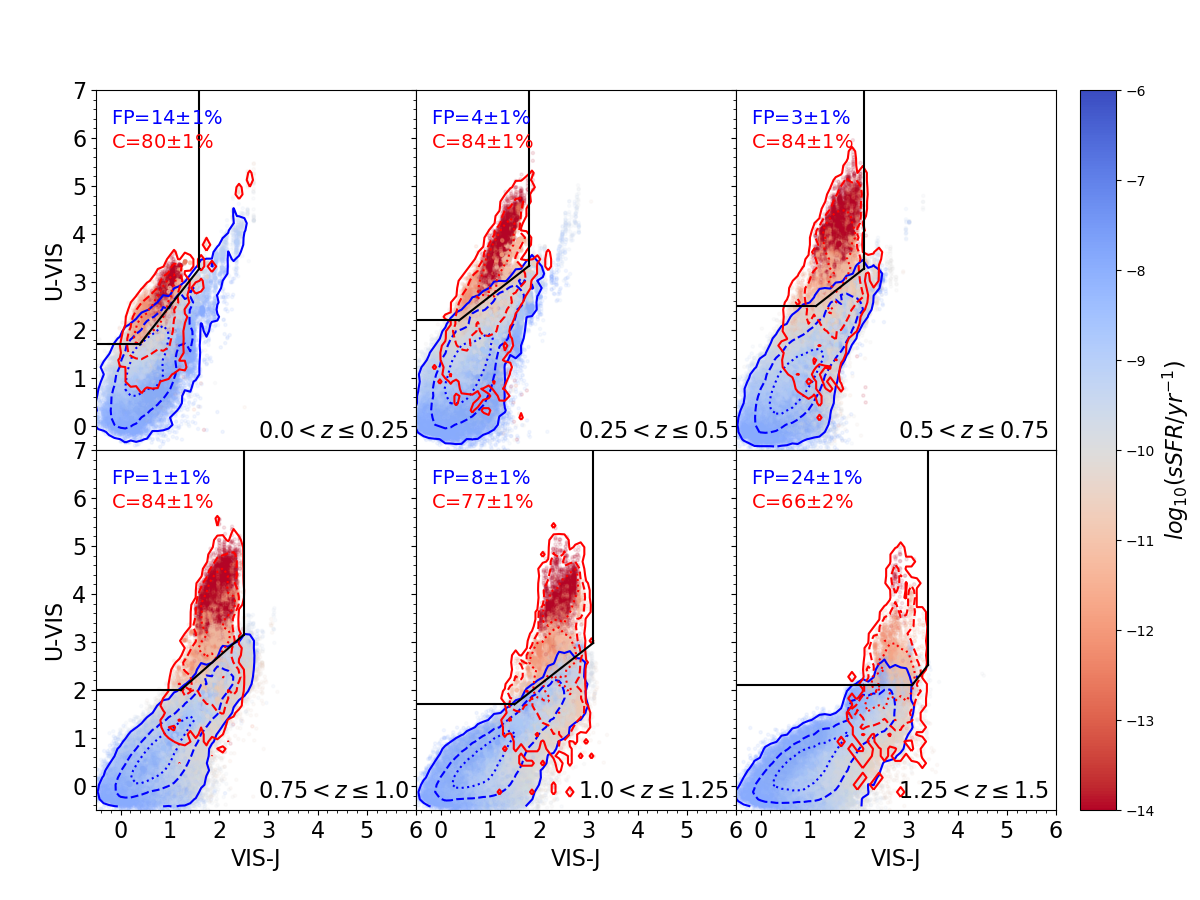}
		\caption{The $(u-VIS)$ vs. $(VIS-J)$ colours obtained from the data set \textit{SED Deep}. Data are shown at different redshifts, from z=0 (\textit{top left}) to z=1.5 (\textit{bottom right}). Galaxies are colour coded depending on their original sSFR. The blue and red lines show the 99.7$\%$ (\textit{solid lines}), 95$\%$ (\textit{dashed lines}) and 68$\%$ (\textit{dotted lines}) contours of the number density of star-forming [$\logten({\rm sSFR/\,yr^{-1}})>-10.5$] and quiescent galaxies [$\logten({\rm sSFR/\,yr^{-1}})<-10.5$], respectively. On the top left of each panel we report the completeness (C) and false-positive fraction (FP) of the quiescent galaxy selection with the corresponding Poisson errors. The black lines show the separation between quiescent and star-forming galaxies that maximises the quantity ${\rm C}\,(1-{\rm FP})$. The selection works well up to at least redshift z=1. }
		\label{fig:Obs_col_UVISJ}
	\end{figure*}
	
	In \autoref{fig:Obs_col_UVISJ} we show the $(u-VIS)$ vs. $(VIS-J)$ colours up to redshift $z=1.5$. We stop our tests at this redshift, because at higher redshifts quiescent galaxies are not detected in the $u$-band in sufficient numbers at the nominal expected depth of the data. Therefore, other techniques will need to be used at higher redshifts. We remind the reader that using  the $u$-band limits our sample significantly: even at lower redshifts, the sub-sample of galaxies visible in the $u$-band in the \Euclid Wide Survey is biased to higher stellar mass galaxies, as explained in \autoref{sec:colours}.  
	Furthermore, the sample of quiescent galaxies detected in the $u$-band is substantially limited in the Euclid Flagship mock galaxy catalogue, so we only consider colours derived from real galaxy observations. \par
	
	We show colours that are determined from the best SED templates, however, we note that colours obtained interpolating the original COSMOS2015 fluxes show a similar behaviour, and the analysis using these provide compatible results (see \autoref{tab:UVISJ_criteria}). The results of the $Flag$ data sets, which we report only for completeness, and we do not use further in the analysis, are consistent with the ones derived using the SED data sets. To simulate photometric errors, we randomly scatter the fluxes of all bands, with a scatter that depends on the expected survey noise (see \autoref{sec:colours_rg}). 
	\par 
	
	Quiescent and star-forming galaxies show some evolution with redshift in both $(u-VIS)$ and $(VIS-J)$ colours. This is expected, since the filters trace different parts of the galaxy spectra at different redshifts, and also the best fitting galaxy templates evolve with redshift. Similarly to the $UVJ$ colour selection, we describe the area occupied by quiescent galaxies at each redshift (black solid lines) as:
	
	\begin{equation}
	\begin{split}
	& (u-VIS) > m \, (VIS-J) +q\,, \\
	& (u-VIS) > L_{\rm low}\,,\, {\rm and} \\
	& (VIS-J) < L_{\rm up}\,. \\
	\end{split}\label{eq:eq_UVISJ}
	\end{equation}
	
	\noindent Considering this description, we derive the best line to isolate quiescent galaxies by maximising the quantity ${\rm C}\,(1-{\rm FP})$. C is the completeness, i.e., the fraction of true quiescent galaxies [$\logten({\rm sSFR/\,yr^{-1}})<$-10.5] that are within the selection, and FP is the false-positive fraction, i.e., the fraction of star-forming galaxies [$\logten({\rm sSFR/\,yr^{-1}})>$-10.5] in the sample lying  within the selection. We decide to maximise the quantity ${\rm C}\,(1-{\rm FP})$ because, generally, the criteria that maximises the completeness corresponds to a false-positive fraction higher than the completeness, whereas the criteria that minimise the false-positive fraction corresponds to a very low completeness. 
	The best separation criterion is derived comparing all lines described by parameters within the intervals of $m\in[0,2[$, $q\in[-2,3[$, $L_{\rm low}\in[0,3[$, and $L_{\rm up}\in[0,4[$ and considering a step of 0.1 for all parameters. \par
	
	We repeat the procedure for the data sets obtained from real galaxy observations (data sets \textit{SED} and \textit{Int}).
	All values derived for each data set are presented in \autoref{tab:UVISJ_criteria}. We then combine the results by averaging the completeness and false-positive fraction of all data sets in the considered parameter space and we derive the best line of separation for quiescent galaxies by maximising again the quantity $\overline{\rm C}\, (1-\overline{\rm FP})$. 
	Note that we do not average the best lines of each data set; we average the completeness and false positive fraction of each possible line in the four data sets and \textit{then} derive the best line. Moreover, we do not apply any weight on the different data sets, as each of them has different drawbacks and strong points. For example, the $SED$ data sets have photometric errors similar to what is expected for \Euclid, but the $Int$ data do not apriori assume a shape for the SED.  \par
	
	\begin{figure}
		\centering
		\includegraphics[trim= 0 20 0 0, clip,width=1\linewidth, keepaspectratio]{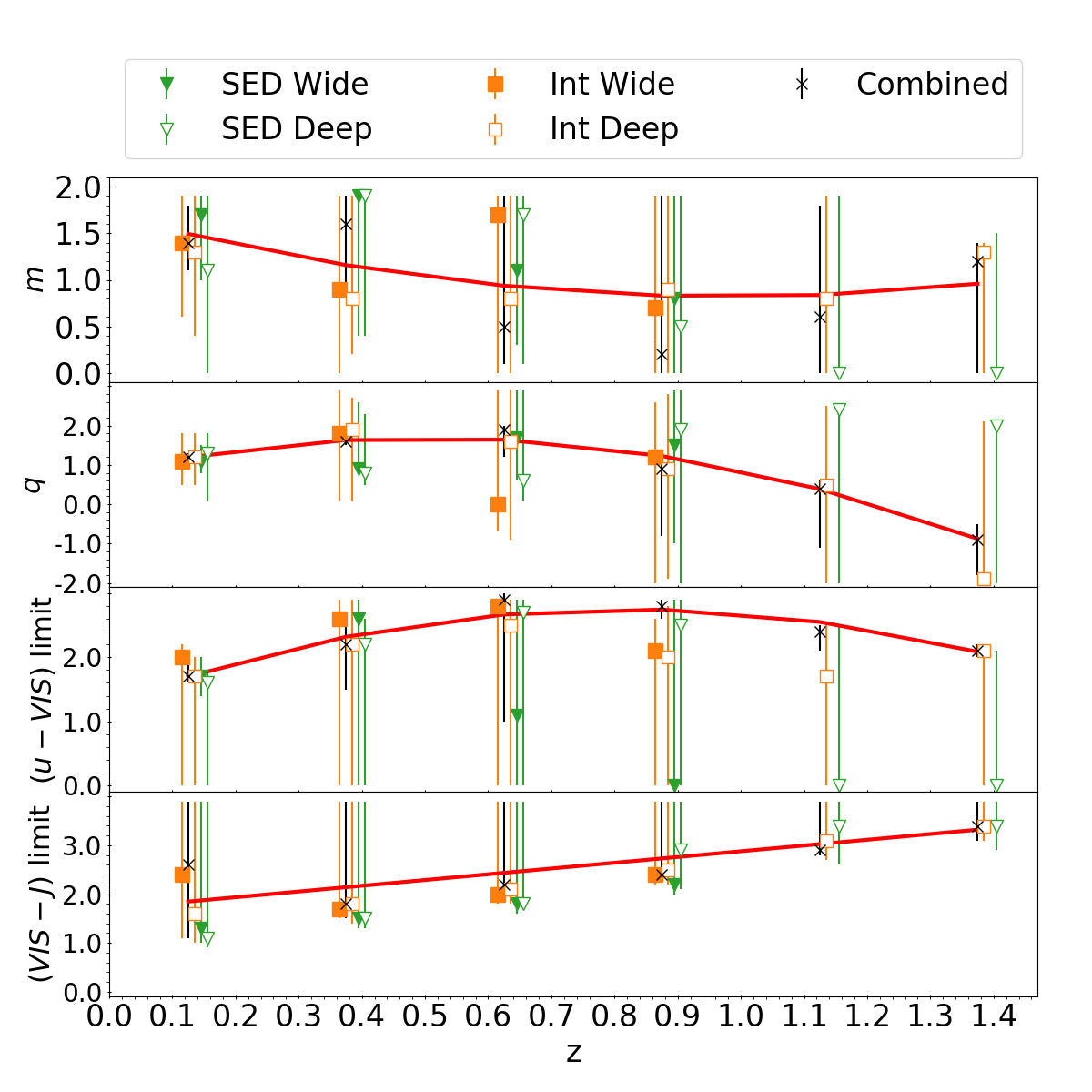}
		\caption{Redshift evolution of the parameters in \autoref{eq:eq_UVISJ} that describes the area isolating quiescent galaxies.
			\textit{From top to bottom:} the slope, the intercept, the lower limit in $(u-VIS)$ colours and the upper limits in the $(VIS-J)$ colours.
			Mock observations are obtained from the best fitting SED template describing the COSMOS2015 observations (\textit{orange squares}) and from the interpolation of the COSMOS2015 observations (\textit{green triangles}). 
			We consider the observational depth planned for both the \Euclid Wide Survey (\textit{filled symbols}) and the \Euclid Deep Survey (\textit{empty symbols}). \textit{Black crosses} correspond to the best-line derived considering the average completeness and false-positive fraction for the four data sets. Coloured data points are slightly shifted horizontally for clarity, while black crosses mark the centre of each bin.
			The red solid line shows the best fit for each parameter (see  \autoref{eq:ev_UVISJ_z}), as derived from the average completeness and false-positive fraction. Marginalised error bars correspond to the parameters values for which the quantity ${\rm C}\,(1-{\rm FP})$ varies by less than 10$\%$ in each different data set.}
		\label{fig:uVISJ_zevol}
	\end{figure}
	
	\begin{table}
		\caption{Best selection criteria for the $(u-VIS)$ and $(VIS-J)$ observed colours at different redshifts, as described in \autoref{eq:eq_UVISJ}.  The last two columns report the completeness (C) and false-positive fraction (FP) of each selection.} 
		\centering 
		\resizebox{0.49\textwidth}{!}{
		\begin{tabular}{c c |c c c c| c c}
			\hline\hline 
			data set & $\langle z\rangle $ & $m$ & $q$ & $L_{\rm low}$ & $L_{\rm up}$ & C & FP\\
			\hline
			& 0.125 & 1.4 & 1.1 & 2.0 & 2.4 & 74$\pm$1$\%$ & 15$\pm$1$\%$\\
			SED & 0.375 & 0.9 & 1.8 & 2.6 & 1.7 & 92$\pm$1$\%$ & 3$\pm$1$\%$\\
			Wide & 0.625 & 1.7 & 0.0 & 2.8 & 2.0 & 84$\pm$1$\%$ & 3$\pm$1$\%$\\
			& 0.875 & 0.7 & 1.2 & 2.1 & 2.4 & 79$\pm$1$\%$ & 5$\pm$1$\%$\\
			\hline
			    & 0.125 & 1.3 & 1.2 & 1.7 & 1.6 & 80$\pm$1$\%$ & 14$\pm$1$\%$ \\
			    & 0.375 & 0.8 & 1.9 & 2.2 & 1.8 & 84$\pm$1$\%$ & 4$\pm$1$\%$\\
			SED & 0.625 & 0.8 & 1.6 & 2.5 & 2.1 & 84$\pm$1$\%$ & 3$\pm$1$\%$\\
			Deep& 0.875 & 0.9 & 0.9 & 2.0 & 2.5 & 84$\pm$1$\%$ & 1$\pm$1$\%$\\
			    & 1.125 & 0.8 & 0.5 & 1.7 & 3.1 & 77$\pm$1$\%$ & 8$\pm$1$\%$ \\
			    & 1.375 & 1.3 & -1.9 & 2.1 & 3.4 & 66$\pm$2$\%$ & 24$\pm$1$\%$\\
			\hline
			& 0.125 & 1.7 & 1.1 & 1.7 & 1.3 & 63$\pm3\%$ & 19$\pm1\%$\\
			Int & 0.375 & 1.9 & 0.9 & 2.6 & 1.5 & 91$\pm3\%$ & 11$\pm1\%$\\
			Wide & 0.625 & 1.1 & 1.7 & 1.1 & 1.8 & 83$\pm4\%$ & 12$\pm1\%$\\
			& 0.875 & 0.8 & 1.5 & 0.0 & 2.2 & 72$\pm5\%$ & 18$\pm2\%$\\
			\hline
			& 0.125 & 1.1 & 1.3 & 1.6 & 1.1 & 40$\pm1\%$ & 21$\pm1\%$\\
			& 0.375 & 1.9 & 0.8 & 2.2 & 1.5 & 39$\pm1\%$ & 12$\pm1\%$\\
			Int & 0.625 & 1.7 & 0.6 & 2.7 & 1.8 & 54$\pm1\%$ & 12$\pm1\%$\\
			Deep & 0.875 & 0.5 & 1.9 & 2.5 & 2.9 & 61$\pm1\%$ & 15$\pm1\%$\\
			& 1.125 & 0.0 & 2.4 & 0.0 & 3.4 & 55$\pm2\%$ & 15$\pm1\%$\\
			& 1.375 & 0.0 & 2.0 & 0.0 & 3.4 & 49$\pm2\%$ & 23$\pm1\%$\\
			\hline
			Flag & 0.125 & 1.3 & 1.0 & 2.0 & 1.6 & 95$\pm$9$\%$ & 0$\pm$1$\%$\\
			Wide$^a$ & 0.375 & 0.9 & 1.5 & 2.8 & 1.6 & 94$\pm$17$\%$ & 2$\pm$2$\%$\\
			     & 0.625 & 0.5 & 1.9 & 0.0 & 1.9 & 60$\pm$20$\%$ & 16$\pm$9$\%$\\
			\hline
			     & 0.125 & 1.4 & 0.9 & 2.1 & 1.6 & 97$\pm$8$\%$ & 3$\pm$1$\%$\\
			Flag & 0.375 & 1.8 & 0.4 & 2.7 & 1.9 & 87$\pm$7$\%$ & 3$\pm$1$\%$\\
			Deep$^a$ & 0.625 & 0.0 & 2.9 & 0.0 & 2.2 & 77$\pm$6$\%$ & 15$\pm$2$\%$\\
			     & 0.875 & 0.1 & 2.3 & 0.0 & 2.4 & 62$\pm$8$\%$ & 19$\pm$4$\%$ \\

2		\end{tabular}}
    $^a$This data set is not used to derive the final colour selection as it is not big enough for statistical purposes.
    	\label{tab:UVISJ_criteria}
	\end{table}
	
	In order to provide galaxy selection criteria at different redshifts, we derive the redshift evolution of each parameter in \autoref{eq:eq_UVISJ}. This is done from the average completeness and false-positive fraction to ensure the stability of the final results compared to the method used to obtain mock observations. Because the errors of the parameters are correlated, we cannot perform an independent fit to the evolution of the parameters that describe the selection area. To bypass this issue, we therefore derive the evolution of each parameter in a sequential order. In particular, we start by extracting the redshift evolution of the slope ($m$) by considering the slope value that simultaneously maximises the average completeness and minimises the average false-positive fraction. In the fit we include the marginalised errors obtained by selecting all slopes that result to $\overline{\rm C}\,(1-\overline{\rm FP})>0.975\, {\rm max}[\overline{\rm C}\,(1-\overline{\rm FP})]$. This corresponds to a maximum error of 10$\%$ of the $C\,(1-{\rm FP})$ of any single data set. Second, we derive the redshift evolution of the intercept $q$, considering all lines that satisfy the same $\overline{\rm C}\,(1-\overline{\rm FP})$ selection, but in addition have slope values equal to the ones predicted with the slope-redshift evolution. Similarly, we include the derived slope and intercept in the redshift evolution in the fit for the $L_{\rm low}$ redshift evolution, and we include in this the evolution of both the slope ($m$), intercept ($q$), and the $(u-VIS)$ lower limit ($L_{\rm low}$) to derive the redshift evolution of the $(VIS-J)$ upper limit ($L_{\rm up}$). The resulting redshift evolution of each parameter is shown in \autoref{fig:uVISJ_zevol} and is described by:
	\begin{equation}
	\begin{aligned} 
	& m = 0.91\,z^2 - 1.80\,z +1.70\,,   \\
	& q = -3.40\,z^2 + 3.44\,z +0.82\,, \\
	& L_{\rm low} = -2.17\,z^2 + 3.56\,z + 1.29\,,  \\
	& L_{\rm up} =  1.18\,z + 1.70\,.\\
	\end{aligned}\label{eq:ev_UVISJ_z}
	\end{equation}
	The evolution of the $(VIS-J)$ limit ($L_{\rm up}$) is well described by a linear relation, while we consider a quadratic polynomial for the slope $m$, the intercept $q$, and the $(u-VIS)$ limit ($L_{\rm low}$). The completeness and the false-positive fraction do not improve much if we consider higher-order polynomials, while the false-positive fraction increases if we consider lower-order polynomials for the slope $m$ and the $(u-VIS)$ limit.  \par
	
\begin{figure*}
		\centering
		\includegraphics[width=0.48\linewidth, keepaspectratio]{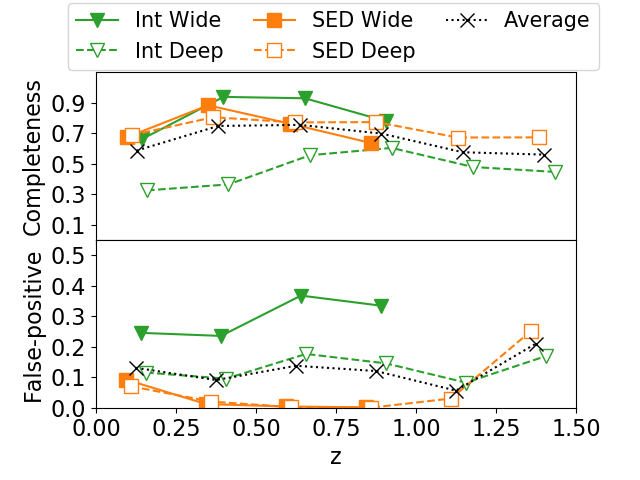}
		\includegraphics[width=0.48\linewidth, keepaspectratio]{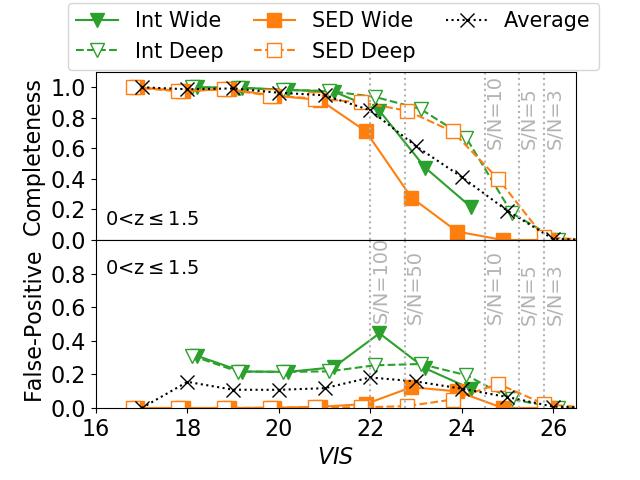}
		\caption{Evolution of the completeness and false-positive fraction with the redshift (\textit{left}) and with the observed $VIS$ magnitude (\textit{right}). Quiescent galaxies are derived considering the best line separation in the $(u-VIS)$ vs. $(VIS-J)$ plane, as described in \autoref{eq:ev_UVISJ_z}. The fractions correspond to the mock observations derived from the best SED template (\textit{orange squares}) and from interpolating the COSMOS2015 observations (\textit{green triangles}), considering the observational depth expected for the \Euclid Wide Survey (\textit{coloured symbols}) and the \Euclid Deep Survey (\textit{empty symbols}). Black crosses are the average values among the four considered data sets. Coloured data points are slightly shifted horizontally for clarity, while black crosses mark the centre of each bin. The grey dotted vertical lines on the right panel show the $VIS$ magnitude corresponding to different S/N cut in the \Euclid Wide Survey.}
		\label{fig:uVISJ_fevol}
	\end{figure*}

	We investigate the accuracy of the selection criteria by calculating the completeness and false-positive fractions in the four data sets derived from real observations (\autoref{fig:uVISJ_fevol}). The average fraction of false-positives is below 15$\%$ at $z\lesssim1.25$, with a maximum of $\sim20\%$ at the highest redshifts. 
	We find that the average completeness is above 55$\%$ at all redshifts. However, the selection is particularly effective at $0.25<z\leq1$,  where the completeness is greater than $\sim70\%$. Note that the completeness of the \textit{Int Deep} data set is quite low. This is due to some galaxies with intermediate colours that are particularly faint and have large photometric errors in the \Euclid Deep Survey and are too faint to be detected in the \Euclid Wide Survey. In general, false-positive fractions are higher for galaxies in the \textit{Int Wide} data set. It is important to consider that both of these data sets are affected by the photometric errors given by the COSMOS2015 catalogues that are typically larger than the errors expected for \Euclid. These inflated photometric errors may have  negatively affected the recovered false-positive fraction and completeness.  \par
	
	In \autoref{fig:uVISJ_fevol} we also show how the completeness and false-positive fraction vary with the observed $VIS$ magnitude for galaxies at $z\leq1.5$. The average false-positive fraction remains almost constant (between 11$\%$ and 16$\%$) for $VIS$ magnitudes between 18 and 25\,mag, with lower false positive fractions for both brighter and fainter objects. On the other hand, a clear trend is visible between the completeness and the $VIS$ observed magnitude, with an average completeness above 80$\%$ at magnitudes brighter than 22\,mag and a steady drop at fainter magnitudes. For both Deep Surveys the drop in completeness happens at  around 23\,mag for both the $Int$ and $SED$ data set. The difference between the completeness in the Wide and Deep Surveys are due to the different uncertainties associated to each galaxy, but also to the different depths in the $u$-band, i.e. the Deep Survey is two magnitudes deeper. At $VIS>22\,$mag, only the bluest quiescent galaxies are detected in the $u$-band. This selection is more important in the Wide Surveys than in the Deep surveys (see also \autoref{fig:VIS_u_sample}). These are galaxies with relatively higher sSFR and are generally the most difficult to disentangle from star-forming galaxies. To give a more quantitative example, galaxies in the $SED$ Wide data set at z$\leq$1.5 and detected in the $H$, $J$, and $u$ filters have a median $\log10(sSFR/{\rm yr^{-1}})=-$12.2. The subsample of galaxies that have the same redshift and detection selection, and also have $VIS>$22\,mag have median $\log10(sSFR/{\rm yr^{-1}})=-$11.1. On the other hand, the same selections in the $SED$ Deep data set produces less of a difference between the two subsamples that have median $\log10(sSFR/{\rm yr^{-1}})=-{\rm11.8}$ and $-{\rm11.7}$, respectively.   \par
	We conclude that the $(u-VIS)$ vs. $(VIS-J)$ colours can be used to isolate quiescent galaxies using the selection described in \autoref{eq:ev_UVISJ_z}, with a generally low contamination by star-forming galaxies and a completeness above 60$\%$, at least up to $z\sim1$. For comparison, the $UVJ$ diagram has been tested and used up to $z\sim3.5$, but, as we previously mentioned, the $U$, $V$, and $J$ rest-frame magnitudes are challenging to derive with only the four \Euclid filters. Indeed, the quiescent galaxy population recovered at z$<$1 with the $UVJ$ diagram with \Euclid has a very low completeness (20$\%$, \autoref{sec:UVJ_Euclid}), making the $(u-VIS)$ and $(VIS-J)$ observed colours the preferred alternative.  This type of analysis will be important and critical when examining the large 15\,000 deg$^{2}$ \Euclid survey area where automation and simplicity will be critical.  
	
	\subsubsection{Redshift separation: the $(VIS-Y)$ vs. $(J-H)$ colours}
	
\begin{figure}
		\centering
		\includegraphics[width=1\linewidth, keepaspectratio]{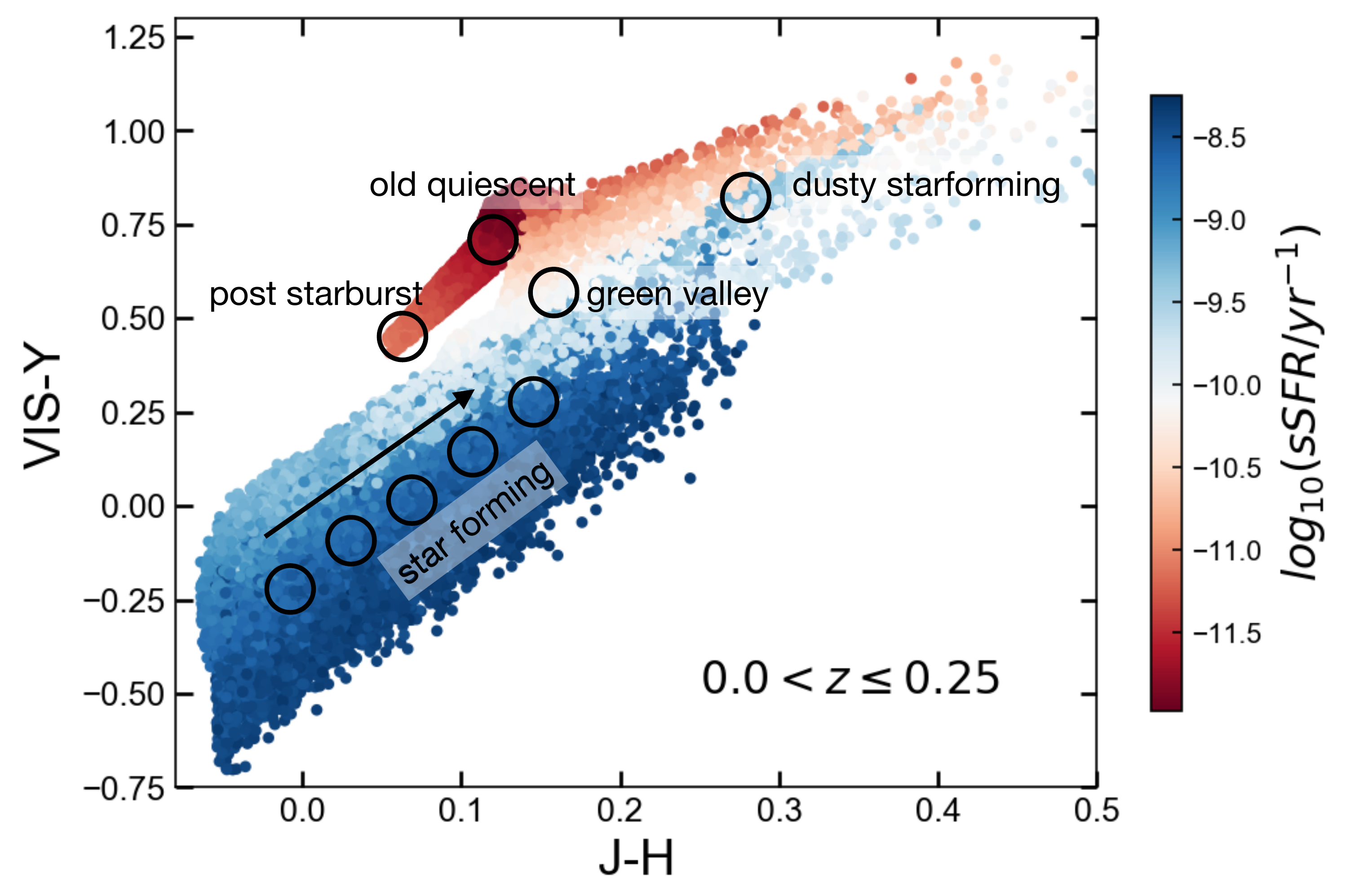} 
		\caption{Colour-colour diagram using simulated \Euclid bands from the Euclid Flagship mock galaxy catalogue in the lowest redshift bin and without observational errors. Galaxies are colour coded depending on their sSFR. The expected colours of some galaxy populations are pin-pointed with black circles.}
		\label{fig:flgship_VISYJH}
	\end{figure}

	We now investigate whether a redshift separation is possible using only the four bands available to \Euclid. We use the $(VIS-Y)$ and $(J-H)$ colours only, which we previously identified as our best case scenario (\autoref{fig:Obs_col}, \autoref{tab:col_mixing}). 
	An idealised case of galaxies in the nearby Universe is shown in \autoref{fig:flgship_VISYJH} in which we plot \Euclid observed colours $(VIS-Y)$ vs. $(J-H)$ from the Euclid Flagship mock galaxy catalogue in the lowest redshift bin, with no addition of photometric errors. 
	Different galaxy populations are indicated by circles and show idealised trends of an evolving galaxy in this colour-colour space. Star-forming galaxies are expected to have blue $(VIS-Y)$ and $(J-H)$ colours, before steadily moving to redder colours as they decrease their star-formation activity and the amount of dust in these systems increases, with a clear separation between quiescent galaxies and dusty star-forming systems. \par
	
\begin{figure*}
		\centering
		\includegraphics[width=1\linewidth, keepaspectratio]{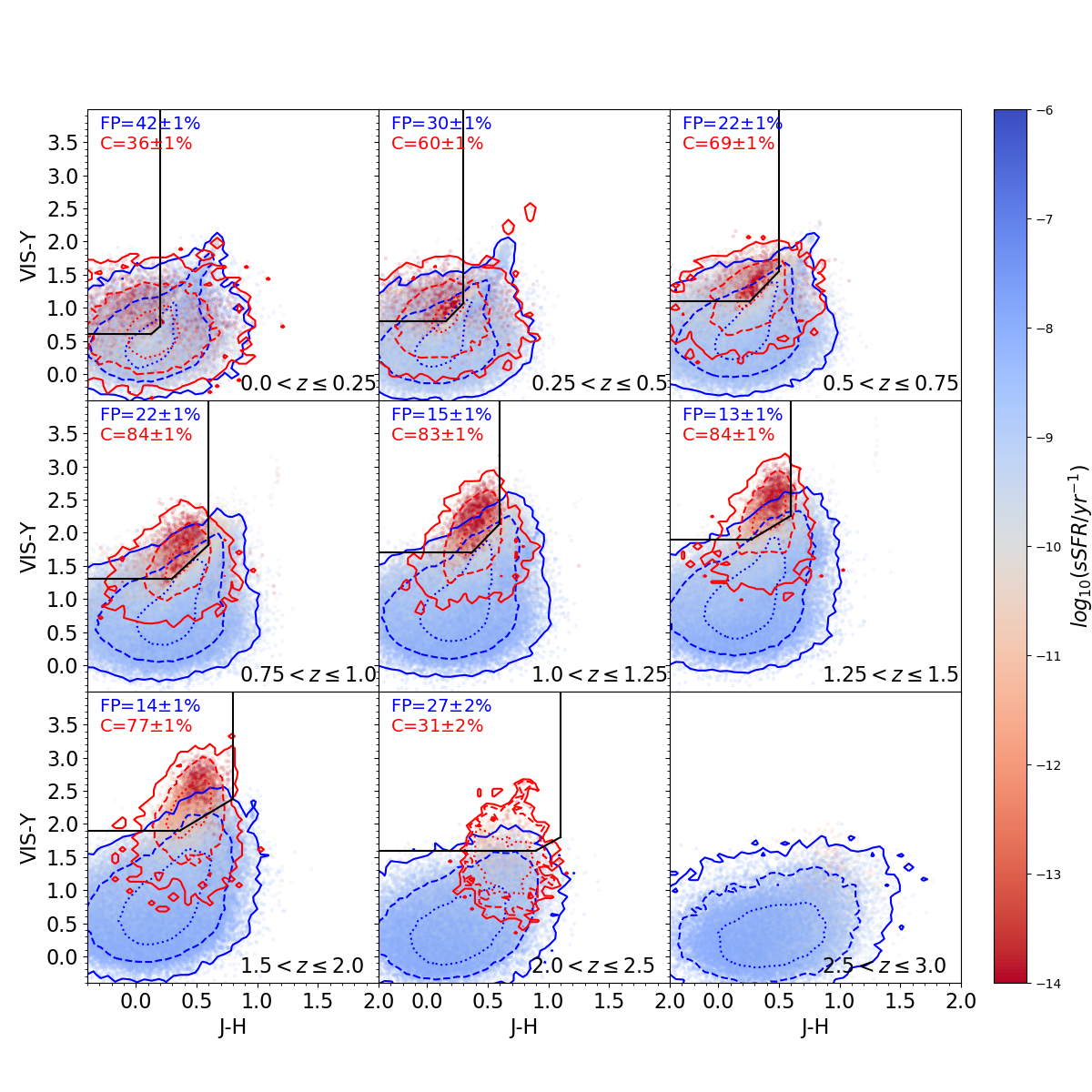}
		\caption{The observed $(VIS-Y)$ vs. $(J-H)$ colours obtained from the  data set \textit{SED Wide}. Data are shown at different redshifts, from z=0 (\textit{top left}) to z=3 (\textit{bottom right}). Galaxies are colour-coded depending on their original sSFR. The blue and red lines show the 99.7$\%$ (\textit{solid lines}), 95$\%$ (\textit{dashed lines}) and 68$\%$ (\textit{dotted lines}) contours of the number density of star-forming [$\logten({\rm sSFR/\,yr^{-1}})>-10.5$] and quiescent galaxies [$\logten({\rm sSFR/\,yr^{-1}})<-10.5$], respectively.  The black lines show the separation between quiescent and star-forming galaxies that maximises the quantity ${\rm C}\,(1-{\rm FP})$.  On the top left of each panel we report the completeness (C) and false-positive fraction (FP) of the quiescent galaxy selection with the corresponding Poisson errors. The selection using \Euclid filters works best in the redshift range  $1<z<2$, where we find a completeness above 65\%.}
		\label{fig:Obs_col_VISYJH}
	\end{figure*}

	Moving away from this idealised case, the inclusion of photometric errors as well as redshift evolution makes the selection of quiescent galaxies more challenging, as shown in \autoref{fig:Obs_col_VISYJH}. We show the selection up to $z=3$, because only a few quiescent galaxies are present in our data sets at higher redshifts. Indeed, if we consider their small number and their mixing in colour space, we realise that the separation criteria would be poorly constrained at higher redshifts.  Colours are shown for the data set \textit{SED Wide} and they are overall similar to the colours of the other five data sets. \par 
	
	We overall find that the star-forming and quiescent galaxies show similar $(VIS-Y)$ and $(J-H)$ colours at low redshift and their separation becomes clearer and cleaner with increasing redshift. This is mainly due to the absence of filters tracing the $\lambda=4000$\AA-break at $z<1$, which is the most prominent feature of an old stellar population.\footnote{To get a sense of which part of the SED is traced by each \Euclid filter at different redshifts, we refer to \autoref{fig:filters}. The red line and open circles shown in the figure represent the observed wavelengths of the 4000\AA-break at different redshifts and over \Euclid's wavelength coverage.} This is not surprising given that the science goals of the \Euclid mission focus their attention at $z>1$. At $z>1$, the $VIS$ band starts to trace near-UV to optical light, while all other bands still trace wavelengths redward of the 4000\AA-break and, indeed, quiescent galaxies have redder $(VIS-Y)$ colours than star-forming objects. At $2<z<3$ the separation is difficult again, as both the $VIS$ and $Y$ filters trace rest-frame $\lambda<4000\,$\AA, while the $J$ and $H$ filters trace rest-frame $\lambda>4000\,$\AA. \par
	
	As in the previous section, we define the area in $VIS$, $Y$, $J$, $H$ colour-space used to select quiescent galaxies as:
	\begin{equation}
	\begin{split}
	& (VIS-Y) > m \, (J-H) +q \,,\\
	& (VIS-Y) >L_{\rm low} \,,\,{\rm and}\\
	& (J-H) < L_{\rm up} \,. \\
	\end{split}\label{eq:eq_VISYJH}
	\end{equation}
	
	Similar to the previous analysis, we derive the best line which separates quiescent and star-forming galaxies by maximising the quantity ${\rm C}\,(1-{\rm FP})$, where C is the completeness and FP is the false-positive fraction. The separation criterion is derived comparing all lines described by parameters inside the intervals of 
	$m\in[0,2[$, $q\in[-2,3[$, $L_{\rm low}\in[0,3[$, and $L_{\rm up}\in[0,2[$ and considering a step of 0.1 for each parameter.
	\par
	A high false-positive fraction, above 30$\%$ at $z<0.5$, and a low completeness, below 70$\%$ at $z<0.75$ reflects the fact that quiescent galaxies are difficult to isolate at low redshifts. For this reason we exclude redshifts below 0.75 when analysing the redshift evolution of the selection area. Results for all six data sets are listed in \autoref{tab:VISYJH_criteria}.  \par 
	
	\begin{table}
		\caption{Best selection criteria for the $(VIS-Y)$ and $(J-H)$ observed colours at different redshifts, as described in \autoref{eq:eq_VISYJH}. The last two columns report the completeness (C) and false-positive fraction (FP) of each selection.} 
		\centering 
		\resizebox{0.49\textwidth}{!}{
		\begin{tabular}{c c |c c c c | c c}
			\hline\hline 
			data set & $\langle z\rangle $ & $m$ & $q$ & $L_{\rm low}$ & $L_{\rm up}$ & C & FP\\
			\hline
			    & 0.125 & 1.6 & 0.4 & 0.6 & 0.2 & 42$\pm$1$\%$ & 36$\pm$1$\%$\\
			    & 0.375 & 1.9 & 0.5 & 0.8 & 0.3 & 60$\pm$1$\%$ & 30$\pm$1$\%$\\
			    & 0.625 & 1.9 & 0.6 & 1.1 & 0.5 & 69$\pm$1$\%$ & 22$\pm$1$\%$\\
			SED & 0.875 & 1.7 & 0.8 & 1.3 & 0.6 & 84$\pm$1$\%$ & 22$\pm$1$\%$\\
			Wide& 1.125 & 1.9 & 1.0 & 1.7 & 0.6 & 83$\pm$1$\%$ & 15$\pm$1$\%$\\
			    & 1.375 & 1.1 & 1.6 & 1.9 & 0.6 & 84$\pm$1$\%$ & 13$\pm$1$\%$\\
			    & 1.750 & 1.1 & 1.5 & 1.9 & 0.8 & 77$\pm$1$\%$ & 14$\pm$1$\%$\\
			    & 2.250 & 1.0 & 0.7 & 1.6 & 1.1 & 31$\pm$2$\%$ & 27$\pm$2$\%$\\
			\hline
			    & 0.125 & 1.7 & 0.4 & 0.6 & 0.3 & 40$\pm$1$\%$ & 28$\pm$1$\%$\\
			    & 0.375 & 1.8 & 0.5 & 0.7 & 0.3 & 53$\pm$1$\%$ & 23$\pm$1$\%$\\
			    & 0.625 & 1.8 & 0.6 & 1.0 & 0.5 & 70$\pm$1$\%$ & 20$\pm$1$\%$\\
			SED & 0.875 & 1.7 & 0.8 & 1.3 & 0.6 & 84$\pm$1$\%$ & 16$\pm$1$\%$\\
			Deep& 1.125 & 1.7 & 1.1 & 1.7 & 0.6 & 87$\pm$1$\%$ & 11$\pm$1$\%$\\
			    & 1.375 & 1.7 & 1.4 & 1.9 & 0.7 & 95$\pm$1$\%$ & 6$\pm$1$\%$\\
			    & 1.750 & 1.2 & 1.5 & 1.9 & 0.8 & 87$\pm$1$\%$ & 8$\pm$1$\%$\\
			    & 2.250 & 1.0 & 1.0 & 1.5 & 1.2 & 50$\pm$1$\%$ & 16$\pm$1$\%$\\
		\hline
			& 0.125 & 1.4 & 0.4 & 0.5 & 0.2 & 37$\pm$1$\%$ & 51$\pm$2$\%$\\
			& 0.375 & 1.9 & 0.3 & 0.7 & 0.3 & 60$\pm$2$\%$ & 38$\pm$1$\%$\\
			& 0.625 & 1.7 & 0.5 & 1.0 & 0.4 & 64$\pm$2$\%$ & 28$\pm$1$\%$\\
			Int & 0.875 & 1.7 & 0.7 & 1.3 & 0.5 & 70$\pm$2$\%$& 27$\pm$1$\%$\\
			Wide & 1.125 & 1.6 & 1.1 & 1.6 & 0.6 & 76$\pm$2$\%$ & 18$\pm$1$\%$\\
			& 1.375 & 1.9 & 1.1 & 1.8 & 0.6 & 77$\pm$3$\%$ & 20$\pm$1$\%$\\
			& 1.750 & 1.5 & 1.2 & 1.8 & 0.8 & 71$\pm$3$\%$ & 19$\pm$1$\%$\\
			& 2.250 & 0.2 & 1.6 & 1.7 & 1.0 & 29$\pm$5$\%$ & 21$\pm$4$\%$\\
			\hline
			& 0.125 & 1.6 & 0.4 & 0.5 & 0.2 & 34$\pm$1$\%$ & 52$\pm$1$\%$\\
			& 0.375 & 1.9 & 0.3 & 0.8 & 0.4 & 37$\pm$1$\%$ & 56$\pm$1$\%$\\
			& 0.625 & 1.7 & 0.5 & 1.1 & 0.4 & 38$\pm$1$\%$ & 41$\pm$1$\%$\\
			Int & 0.875 & 1.8 & 0.7 & 1.3 & 0.5 & 60$\pm$1$\%$ & 37$\pm$1$\%$\\
			Deep & 1.125 & 1.6 & 1.1 & 1.7 & 0.6 & 67$\pm$2$\%$ & 22$\pm$1$\%$\\
			& 1.375 & 1.9 & 1.1 & 1.8 & 0.6 & 70$\pm$3$\%$ & 27$\pm$1$\%$\\
			& 1.750 & 1.5 & 1.2 & 1.9 & 0.8 & 53$\pm$2$\%$ & 25$\pm$1$\%$\\
			& 2.250 & 0.3 & 1.6 & 1.9 & 1.5 & 23$\pm$2$\%$ & 31$\pm$3$\%$\\
			\hline
			& 0.125 & 1.6 & 0.5 & 0.8 & 0.3 & 78$\pm$6$\%$ & 16$\pm$2$\%$\\
			Flag & 0.375 & 1.9 & 0.5 & 0.8 & 0.4 & 67$\pm$5$\%$ & 18$\pm$2$\%$\\
			Wide & 0.625 & 0.9 & 0.9 & 1.1 & 0.7 & 71$\pm$4$\%$ & 34$\pm$2$\%$\\
			& 0.875 & 1.4 & 0.9 & 1.4 & 0.8 & 68$\pm$4$\%$ & 29$\pm$2$\%$\\
			& 1.125 & 1.7 & 1.1 & 1.7 & 0.6 & 64$\pm$8$\%$ & 25$\pm$4$\%$ \\
			\hline
			& 0.125 & 0.9 & 0.7 & 0.8 & 0.3 & 77$\pm$6$\%$ & 36$\pm$4$\%$\\
			Flag & 0.375 & 1.7 & 0.6 & 1.0 & 0.4 & 49$\pm$4$\%$ & 37$\pm$3$\%$\\
			Deep & 0.625 & 0.7 & 1.0 & 1.2 & 0.7 & 63$\pm$3$\%$ & 30$\pm$2$\%$\\
			& 0.875 & 1.4 & 0.9 & 1.4 & 0.8 & 72$\pm$4$\%$ & 25$\pm$2$\%$\\
			& 1.125 & 1.6 & 1.2 & 1.7 & 0.7 & 65$\pm$6$\%$ & 17$\pm$3$\%$
		\end{tabular}}
		\label{tab:VISYJH_criteria}
	\end{table}
	
\begin{figure}
		\centering
		\includegraphics[trim=0 20 0 0,clip,width=1\linewidth, keepaspectratio]{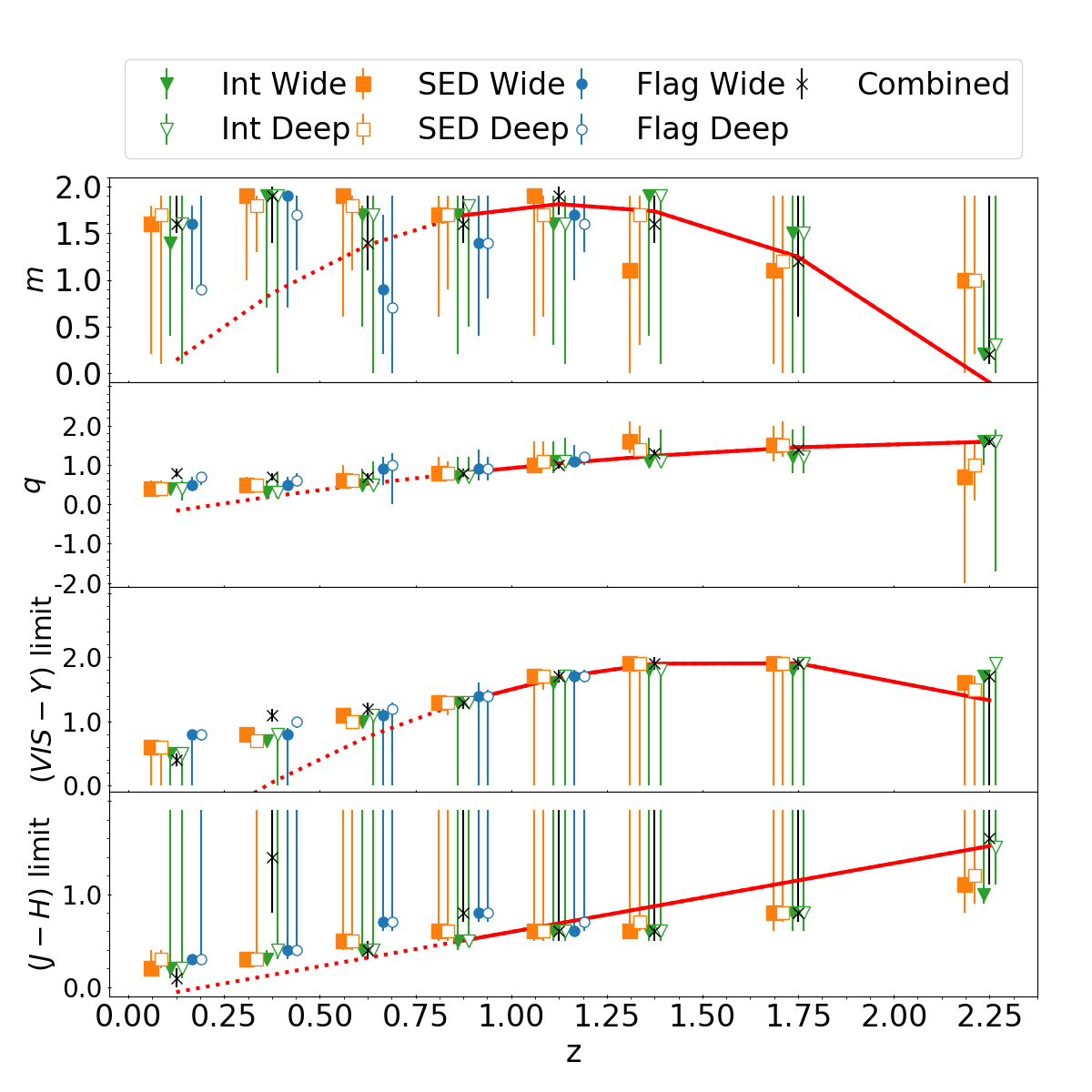}
		\caption{Redshift evolution of the parameters in \autoref{eq:eq_VISYJH} that describe the area isolating quiescent galaxies. \textit{From top to bottom:} the slope, the intercept, the lower limit in $(VIS-Y)$ colours and the upper limits in the $(J-H)$ colours. Mock observations are obtained from the best fitting SED template describing the COSMOS2015 observations (\textit{orange squares}), from the interpolation of the COSMOS2015 observations (\textit{green triangles}) and from the Euclid Flagship mock galaxy catalogue (\textit{blue circles}). We consider the observational depth planned for both the \Euclid Wide Survey (\textit{filled symbols}) and the \Euclid Deep Survey (\textit{empty symbols}). \textit{Black crosses} correspond to the best-line derived considering the average completeness and false-positive fraction for the six data sets. Coloured data points are slightly shifted horizontally for clarity, while black crosses mark the centre of each bin. The red continuous lines show the best fit to the considered points at z>0.75 (see \autoref{eq:ev_VISYJ_z}), as derived from the average completeness and false-positive fraction. The dashed lines show the extrapolation at low redshifts. Marginalised error bars correspond to the parameters values for which the quantity ${\rm C}\,(1-{\rm FP})$ varies by less than 10$\%$ in each different data set.}
		\label{fig:VISYJ_zevol}
	\end{figure}

	We average the results of the mock galaxies of all the six data sets to obtain the evolution of the line separating star-forming and quiescent galaxies with redshift (\autoref{fig:VISYJ_zevol}). In the Euclid Flagship mock galaxy catalogue used for the \textit{Flag Wide} and \textit{Flag Deep} data sets, there are almost no quiescent galaxies at $z>1.25$, but at lower redshift the line separation overall agrees with the value derived from the COSMOS2015 catalogue. As we did for the $(u-VIS)$ and $(VIS-J)$ colours, we adopt a sequential approach that starts from the fit of the slope-redshift evolution, and then uses the results of this fit to derive the redshift evolution of the intercept $q$. The same method is then applied to the $(VIS-Y)$ limit and the $(J-H)$ limit. In the fit of the redshift evolution of each parameter we include marginalised errors obtained by considering all selection criteria with $\overline{\rm C}\,(1-\overline{\rm FP})>0.983\, {\rm max}[\overline{\rm C}\,(1-\overline{\rm FP})]$, which correspond to a maximum error of 10$\%$ in the ${\rm C}\,(1-{\rm FP})$ value of any single data set. Differences in the marginalised error estimates with the $(u-VIS)$ vs. $(VIS-J)$ analysis are due to the different number of data sets considered. By combining the results of the different data sets, the line separating quiescent and star-forming galaxies can be described as a function of redshift as:
	\begin{equation}
	\begin{aligned} 
	& m = -1.59 \, z^2 + 3.66 \, z - 0.30 \,, \\
	& q = -0.33\, z^2 +1.61\, z - 0.36 \,,\\
	& L_{\rm low} = -1.34 \, z^2 + 4.20 \, z - 1.34\,,\\
	& L_{\rm up} =  0.74 \, z - 0.14\,.\\
	\end{aligned}\label{eq:ev_VISYJ_z}
	\end{equation}

	We consider a second-degree polynomial for fitting the slope $m$, the intercept $q$, and the $(VIS-Y)$ limit (C$_{\rm low}$)  and a linear regression for the $(J-H)$ limit (C$_{\rm up}$) . By considering higher-order polynomials the completeness and false-positive fractions at $0.75<z<2.5$ do not change considerably. At the same time, considering lower-order polynomials decreases the average completeness below 50$\%$ and increases the average false-positive fractions above 50$\%$.
	
		\begin{figure*}
		\centering
		\includegraphics[width=0.48\linewidth, keepaspectratio]{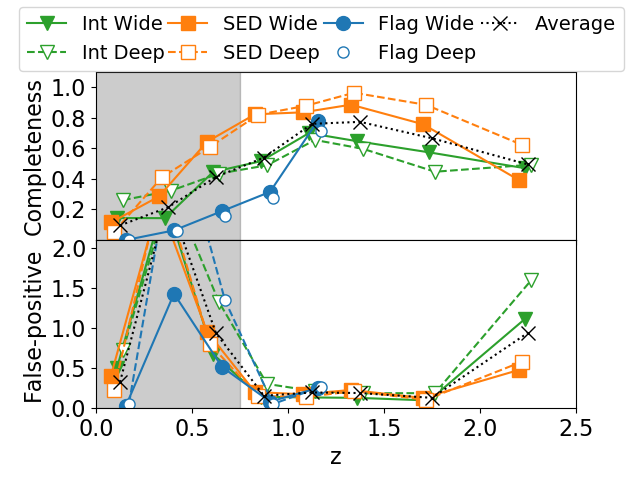}
		\includegraphics[width=0.48\linewidth, keepaspectratio]{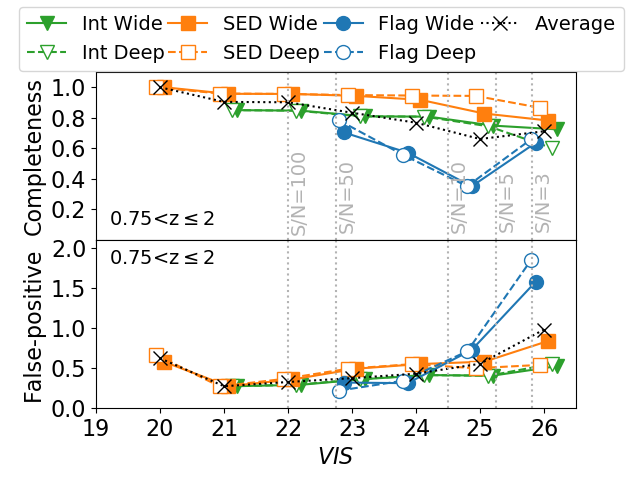}
		\caption{Evolution of the completeness and false-positive fraction with redshift (\textit{left}) and observed $VIS$ magnitude  (\textit{right}). Quiescent galaxies are derived considering the best line separation in the $(VIS-Y)$ vs. $(J-H)$ plane, as described in \autoref{eq:ev_VISYJ_z}. The fractions correspond to mock observations derived from the best SED template (\textit{orange squares}), from interpolating the COSMOS2015 observations (\textit{green triangles}) and from the Euclid Flagship mock galaxy catalogue (\textit{blue circles}). We include results derived considering the observational depth expected for the \Euclid Wide Survey (\textit{coloured symbols}) and the \Euclid Deep Survey (\textit{empty symbols}). Black crosses are the average values among the six considered data sets.  Coloured data points are slightly shifted horizontally for clarity, while black crosses mark the centre of each bin. The grey areas are outside the redshift range used to derive the evolution of the quiescent galaxy selection criteria. The grey dotted vertical lines on the right panel show the $VIS$ magnitude corresponding to different S/N cuts in the \Euclid Wide Survey. }
		\label{fig:VISYJH_fevol}
	\end{figure*}
	
	As for the $(u-VIS)$ and $(VIS-J)$ colours, we verify the quality of the selection criteria in all data sets by calculating the completeness and false-positive fraction for the selection criteria using \autoref{eq:ev_VISYJ_z} (\autoref{fig:VISYJH_fevol}). We advise against extrapolating the selection criteria to $z<0.75$, as the star-forming galaxies will have a high contamination. At $z>2$ the combined effect of poor statistical constraints and the absence of colours that include the 4000\AA$\,$-break makes the selection difficult. The best scenario in this case results in a low completeness and a very high false-positive fraction. However, relaxing the selection criterion mainly increases the false-positive fraction, rather than the completeness.\par 
	In \autoref{fig:VISYJH_fevol}, we also show the completeness and false-positive fraction at different observed $VIS$ magnitudes, for galaxies at redshift $0.75<z<2$. Differently from the results for the $(u-VIS)$ and $(VIS-J)$ colours, the completeness for the $(VIS-J)$ and $(Y-H)$ colours shows a mild decrease with increasing $VIS$ magnitude, with average values around 100$\%$ at $VIS=20$\,mag and around 70$\%$ at $VIS=26$\,mag. The false-positive fraction, on the other hand, shows an increase with increasing $VIS$ observed magnitude, with the average values smaller than 50$\%$ only for objects between $VIS=21\,$mag and $VIS=24\,$mag. We do not find substantial differences between the Wide and Deep Surveys. Most differences arise from a variation in the data sets, particularly between the data sets derived from real galaxy observations ($SED$ and $Int$ data sets) and those from the simulated galaxies ($Flag$ data sets). In particular, as investigated in more details in Appendix \ref{sec:mag_data sets}, the $Flag$ data sets have on average galaxies with lower sSFR and fainter VIS magnitudes than the other two data sets. Star-forming galaxies with relatively low sSFR are generally more difficult to separate from the quiescent galaxies and this influences the recovered completeness and false-positive fraction.
	
	Overall, we conclude that $(VIS-Y)$ and $(J-H)$ colours can be used to select quiescent galaxies at $1<z<2$ ($0.75<z<2$) with an average completeness above 65$\%$ (55$\%$) and with false-positive fractions typically below $\sim20\%$. Therefore, this colour combination is complementary in redshift to the $(u-VIS)$ and $(VIS-J)$ colour selection previously analysed and shows a similar completeness, but a slightly larger false-positive fraction, i.e., below 15$\%$ at $0.25<z<1$ for the $(u-VIS)$ and $(VIS-J)$ colours.  We speculate that other criteria, like galaxy morphologies, could be used in tandem with these colours to improve these selections further.

	\section{Summary}\label{sec:conclusions}
	
	
	Colour-colour selections are widely used and well accepted methods in extragalactic astronomy to separate different galaxy population, such as quiescent and star-forming galaxies. 
	Given the limited number of filters in general and the unusually wide visual filter in particular designed for the \Euclid telescope, it is vital to determine a framework astronomers can use for this purpose with the extensive imaging data that will arise from \Euclid. In this paper, we show that \Euclid filters alone are not sufficient to pin down a best fit template to determine the rest-frame colours based on $U$, $V$, and $J$ bands used in standard selections, nor are they adequate to derive specific star-formation rates. We therefore derive \Euclid specific selection criteria for the separation of quiescent and star-forming galaxies using \Euclid observed colours. \par
	
	To do so, we define three different sets of mock \Euclid observations: i) the first interpolates the multi-wavelength observations of galaxies in the COSMOS2015 catalogue; ii) the second uses the best theoretical template describing the multi-wavelength observations of galaxies in the COSMOS2015 catalogue; iii) the third takes galaxy parameters from the Euclid Flagship mock galaxy catalogue. 
	Each data set contains mock observations for \Euclid's visible $VIS$ filter, and the near-infrared filters NISP $Y$, $J$, and $H$.  Data sets i) and ii) also include CFIS/$u$ band observations.  Similar $u$-band data will be available with other overlapping surveys such as LSST. \par
	
	By selecting galaxy types in the commonly accepted $UVJ$ plane derived from these mock observations, we only recover $\sim20\%$ of the original quiescent galaxy population up to redshifts $z = 3$. The reason for this low success rate is the difficulty of deriving accurate $(U-V)$ and $(V-J)$ colours with only four filters as is the case for the \Euclid mission. Even worse, when we use the sSFR derived from the four \Euclid filters to isolate quiescent galaxies, we recover only 9$\%$ of the original quiescent galaxy population. \par
	
	We find that the most effective way to separate quiescent from star-forming galaxies with observed colours is the combination of $(u-VIS)$ and $(VIS-J)$ colours. This filter combination will be available thanks to the \Euclid-specific follow-up ancillary ground-based $u$-band observations. For this colour combination,  the bulk of quiescent and star-forming galaxies (i.e., the areas containing 68$\%$ of the number density of these two classes of galaxies) are completely separated. We derive the quantitative separation of the two galaxy populations by simultaneously maximising the completeness of the quiescent galaxy recovery and minimising the number of false-positives. We further parameterise the evolution of this fitting with redshift. The proposed line allows for a selection of quiescent galaxies (with a recovery of more than 55$\%$ up to $z\sim1$) while keeping the average fraction of false-positive below 15$\%$. We find the highest success rates in the redshift range $0.25<z<1$, where the completeness is above $\sim$70$\%$. \par
	
	We also tested the performance of separating galaxy types when using only the four filters on board the \Euclid telescope. Of the five colour combinations we tested, the $(VIS-Y)$ and $(J-H)$ colours are the most efficient for isolating quiescent galaxies. A drawback lies at low redshifts: due to the absence of strong spectral features inside these filters at $z<0.75$, quiescent and star-forming galaxies have similar colours. We therefore offer selection criteria only for higher redshifts. We do this by maximising the selection completeness and, at the same time, minimising the false-positive fraction. The derived selection criteria allow the user to select a sample of quiescent galaxies at $0.75<z<2$ with average completeness above 55$\%$, and an average false-positive fraction below 20$\%$. The selection works best in the redshift range $1<z<2$, where we find a completeness above 65$\%$.   
	
	\Euclid will provide additional information besides colours, such as the resolved structures of galaxies up to high redshifts. Using a combination of colours and morphologies, we expect that success rates will increase and contamination rates will decrease. Similar improvements could be achieved with the addition of spectroscopic information from the NISP spectra, when available. This will be tested in future work.  \par

	\section*{Acknowledgements}
	UK acknowledges support from STFC. CC acknowledges the support of the  STFC Cosmic Vision funding. CT acknowledges funding from the INAF PRIN-SKA 2017 program 1.05.01.88.04.
	\AckEC
	\bibliographystyle{mnras}
	\bibliography{Euclidcolours_v2.1}	

\begin{thebibliography}{}
\makeatletter
\relax
\def\mn@urlcharsother{\let\do\@makeother \do\$\do\&\do\#\do\^\do\_\do\%\do\~}
\def\mn@doi{\begingroup\mn@urlcharsother \@ifnextchar [ {\mn@doi@}
  {\mn@doi@[]}}
\def\mn@doi@[#1]#2{\def\@tempa{#1}\ifx\@tempa\@empty \href
  {http://dx.doi.org/#2} {doi:#2}\else \href {http://dx.doi.org/#2} {#1}\fi
  \endgroup}
\def\mn@eprint#1#2{\mn@eprint@#1:#2::\@nil}
\def\mn@eprint@arXiv#1{\href {http://arxiv.org/abs/#1} {{\tt arXiv:#1}}}
\def\mn@eprint@dblp#1{\href {http://dblp.uni-trier.de/rec/bibtex/#1.xml}
  {dblp:#1}}
\def\mn@eprint@#1:#2:#3:#4\@nil{\def\@tempa {#1}\def\@tempb {#2}\def\@tempc
  {#3}\ifx \@tempc \@empty \let \@tempc \@tempb \let \@tempb \@tempa \fi \ifx
  \@tempb \@empty \def\@tempb {arXiv}\fi \@ifundefined
  {mn@eprint@\@tempb}{\@tempb:\@tempc}{\expandafter \expandafter \csname
  mn@eprint@\@tempb\endcsname \expandafter{\@tempc}}}

\bibitem[\protect\citeauthoryear{{Allen}}{{Allen}}{1976}]{Allen1976}
{Allen} C.~W.,  1976, {Astrophysical Quantities, London: Athlone (3rd edition)}

\bibitem[\protect\citeauthoryear{{Andreon}}{{Andreon}}{2018}]{Andreon2018}
{Andreon} S.,  2018, \mn@doi [\aap] {10.1051/0004-6361/201832627}, \href
  {https://ui.adsabs.harvard.edu/abs/2018A&A...617A..53A} {617, A53}

\bibitem[\protect\citeauthoryear{{Arnouts}, {Cristiani}, {Moscardini},
  {Matarrese}, {Lucchin}, {Fontana}  \& {Giallongo}}{{Arnouts}
  et~al.}{1999}]{Arnouts1999}
{Arnouts} S.,  {Cristiani} S.,  {Moscardini} L.,  {Matarrese} S.,  {Lucchin}
  F.,  {Fontana} A.,   {Giallongo} E.,  1999, \mn@doi [\mnras]
  {10.1046/j.1365-8711.1999.02978.x}, \href
  {http://adsabs.harvard.edu/abs/1999MNRAS.310..540A} {310, 540}

\bibitem[\protect\citeauthoryear{{Baldry}, {Glazebrook}, {Brinkmann},
  {Ivezi{\'c}}, {Lupton}, {Nichol}  \& {Szalay}}{{Baldry}
  et~al.}{2004}]{Baldry2004}
{Baldry} I.~K.,  {Glazebrook} K.,  {Brinkmann} J.,  {Ivezi{\'c}} {\v Z}.,
  {Lupton} R.~H.,  {Nichol} R.~C.,   {Szalay} A.~S.,  2004, \mn@doi [\apj]
  {10.1086/380092}, \href {http://adsabs.harvard.edu/abs/2004ApJ...600..681B}
  {600, 681}

\bibitem[\protect\citeauthoryear{{Bell} et~al.,}{{Bell}
  et~al.}{2004}]{Bell2004}
{Bell} E.~F.,  et~al., 2004, \mn@doi [\apj] {10.1086/420778}, \href
  {http://adsabs.harvard.edu/abs/2004ApJ...608..752B} {608, 752}

\bibitem[\protect\citeauthoryear{{Bisigello}, {Caputi}, {Grogin}  \&
  {Koekemoer}}{{Bisigello} et~al.}{2018}]{Bisigello2018}
{Bisigello} L.,  {Caputi} K.~I.,  {Grogin} N.,   {Koekemoer} A.,  2018, \mn@doi
  [\aap] {10.1051/0004-6361/201731399}, \href
  {https://ui.adsabs.harvard.edu/abs/2018A&A...609A..82B} {609, A82}

\bibitem[\protect\citeauthoryear{{Blanton} et~al.,}{{Blanton}
  et~al.}{2003a}]{Blanton2003}
{Blanton} M.~R.,  et~al., 2003a, \mn@doi [\apj] {10.1086/375776}, \href
  {http://adsabs.harvard.edu/abs/2003ApJ...592..819B} {592, 819}

\bibitem[\protect\citeauthoryear{{Blanton} et~al.,}{{Blanton}
  et~al.}{2003b}]{Blanton2003a}
{Blanton} M.~R.,  et~al., 2003b, \mn@doi [\apj] {10.1086/375528}, \href
  {http://adsabs.harvard.edu/abs/2003ApJ...594..186B} {594, 186}

\bibitem[\protect\citeauthoryear{{Blanton} et~al.,}{{Blanton}
  et~al.}{2005}]{Blanton2005}
{Blanton} M.~R.,  et~al., 2005, \mn@doi [\aj] {10.1086/429803}, \href
  {http://adsabs.harvard.edu/abs/2005AJ....129.2562B} {129, 2562}

\bibitem[\protect\citeauthoryear{{Bruzual} \& {Charlot}}{{Bruzual} \&
  {Charlot}}{2003}]{Bruzual2003}
{Bruzual} G.,  {Charlot} S.,  2003, \mn@doi [\mnras]
  {10.1046/j.1365-8711.2003.06897.x}, \href
  {http://adsabs.harvard.edu/abs/2003MNRAS.344.1000B} {344, 1000}

\bibitem[\protect\citeauthoryear{{Calzetti}, {Armus}, {Bohlin}, {Kinney},
  {Koornneef}  \& {Storchi-Bergmann}}{{Calzetti} et~al.}{2000}]{Calzetti2000}
{Calzetti} D.,  {Armus} L.,  {Bohlin} R.~C.,  {Kinney} A.~L.,  {Koornneef} J.,
   {Storchi-Bergmann} T.,  2000, \mn@doi [\apj] {10.1086/308692}, \href
  {http://adsabs.harvard.edu/abs/2000ApJ...533..682C} {533, 682}

\bibitem[\protect\citeauthoryear{{Carretero}, {Castander}, {Gazta{\~n}aga},
  {Crocce}  \& {Fosalba}}{{Carretero} et~al.}{2015}]{Carretero2015}
{Carretero} J.,  {Castander} F.~J.,  {Gazta{\~n}aga} E.,  {Crocce} M.,
  {Fosalba} P.,  2015, \mn@doi [\mnras] {10.1093/mnras/stu2402}, \href
  {http://adsabs.harvard.edu/abs/2015MNRAS.447..646C} {447, 646}

\bibitem[\protect\citeauthoryear{{Cassata} et~al.,}{{Cassata}
  et~al.}{2007}]{Cassata2007}
{Cassata} P.,  et~al., 2007, \mn@doi [\apjs] {10.1086/516591}, \href
  {https://ui.adsabs.harvard.edu/abs/2007ApJS..172..270C} {172, 270}

\bibitem[\protect\citeauthoryear{{Chabrier}}{{Chabrier}}{2003}]{Chabrier2003}
{Chabrier} G.,  2003, \mn@doi [\pasp] {10.1086/376392}, \href
  {http://adsabs.harvard.edu/abs/2003PASP..115..763C} {115, 763}

\bibitem[\protect\citeauthoryear{{Crocce}, {Castander}, {Gazta{\~n}aga},
  {Fosalba}  \& {Carretero}}{{Crocce} et~al.}{2015}]{Crocce2015}
{Crocce} M.,  {Castander} F.~J.,  {Gazta{\~n}aga} E.,  {Fosalba} P.,
  {Carretero} J.,  2015, \mn@doi [\mnras] {10.1093/mnras/stv1708}, \href
  {http://adsabs.harvard.edu/abs/2015MNRAS.453.1513C} {453, 1513}

\bibitem[\protect\citeauthoryear{{Cropper} et~al.,}{{Cropper}
  et~al.}{2010}]{Cropper2010}
{Cropper} M.,  et~al., 2010. , \mn@doi{10.1117/12.857224}

\bibitem[\protect\citeauthoryear{{Duncan} et~al.,}{{Duncan}
  et~al.}{2014}]{Duncan2014}
{Duncan} K.,  et~al., 2014, \mn@doi [\mnras] {10.1093/mnras/stu1622}, \href
  {https://ui.adsabs.harvard.edu/abs/2014MNRAS.444.2960D} {444, 2960}

\bibitem[\protect\citeauthoryear{{Fang} et~al.,}{{Fang}
  et~al.}{2018}]{Fang2018}
{Fang} J.~J.,  et~al., 2018, \mn@doi [\apj] {10.3847/1538-4357/aabcba}, \href
  {https://ui.adsabs.harvard.edu/abs/2018ApJ...858..100F} {858, 100}

\bibitem[\protect\citeauthoryear{{Fosalba}, {Gazta{\~n}aga}, {Castander}  \&
  {Crocce}}{{Fosalba} et~al.}{2015a}]{Fosalba2015b}
{Fosalba} P.,  {Gazta{\~n}aga} E.,  {Castander} F.~J.,   {Crocce} M.,  2015a,
  \mn@doi [\mnras] {10.1093/mnras/stu2464}, \href
  {http://adsabs.harvard.edu/abs/2015MNRAS.447.1319F} {447, 1319}

\bibitem[\protect\citeauthoryear{{Fosalba}, {Crocce}, {Gazta{\~n}aga}  \&
  {Castander}}{{Fosalba} et~al.}{2015b}]{Fosalba2015a}
{Fosalba} P.,  {Crocce} M.,  {Gazta{\~n}aga} E.,   {Castander} F.~J.,  2015b,
  \mn@doi [\mnras] {10.1093/mnras/stv138}, \href
  {http://adsabs.harvard.edu/abs/2015MNRAS.448.2987F} {448, 2987}

\bibitem[\protect\citeauthoryear{{Fritz} et~al.,}{{Fritz}
  et~al.}{2014}]{Fritz2014}
{Fritz} A.,  et~al., 2014, \mn@doi [\aap] {10.1051/0004-6361/201322379}, \href
  {http://adsabs.harvard.edu/abs/2014A%26A...563A..92F} {563, A92}

\bibitem[\protect\citeauthoryear{{Fumagalli} et~al.,}{{Fumagalli}
  et~al.}{2012}]{Fumagalli2012}
{Fumagalli} M.,  et~al., 2012, \mn@doi [\apjl] {10.1088/2041-8205/757/2/L22},
  \href {https://ui.adsabs.harvard.edu/abs/2012ApJ...757L..22F} {757, L22}

\bibitem[\protect\citeauthoryear{{Giallongo}, {Salimbeni}, {Menci}, {Zamorani},
  {Fontana}, {Dickinson}, {Cristiani}  \& {Pozzetti}}{{Giallongo}
  et~al.}{2005}]{Giallongo2005}
{Giallongo} E.,  {Salimbeni} S.,  {Menci} N.,  {Zamorani} G.,  {Fontana} A.,
  {Dickinson} M.,  {Cristiani} S.,   {Pozzetti} L.,  2005, \mn@doi [\apj]
  {10.1086/427819}, \href
  {https://ui.adsabs.harvard.edu/abs/2005ApJ...622..116G} {622, 116}

\bibitem[\protect\citeauthoryear{{Ibata} et~al.,}{{Ibata}
  et~al.}{2017}]{Ibata2017}
{Ibata} R.~A.,  et~al., 2017, \mn@doi [\apj] {10.3847/1538-4357/aa855c}, \href
  {http://adsabs.harvard.edu/abs/2017ApJ...848..128I} {848, 128}

\bibitem[\protect\citeauthoryear{{Ilbert} et~al.,}{{Ilbert}
  et~al.}{2006}]{Ilbert2006}
{Ilbert} O.,  et~al., 2006, \mn@doi [\aap] {10.1051/0004-6361:20065138}, \href
  {http://adsabs.harvard.edu/abs/2006A%26A...457..841I} {457, 841}

\bibitem[\protect\citeauthoryear{{Ilbert} et~al.,}{{Ilbert}
  et~al.}{2009}]{Ilbert2009}
{Ilbert} O.,  et~al., 2009, \mn@doi [\apj] {10.1088/0004-637X/690/2/1236},
  \href {https://ui.adsabs.harvard.edu/abs/2009ApJ...690.1236I} {690, 1236}

\bibitem[\protect\citeauthoryear{{Ivezic} et~al.,}{{Ivezic}
  et~al.}{2008}]{Ivezic2008}
{Ivezic} Z.,  et~al., 2008, \mn@doi [Serbian Astronomical Journal]
  {10.2298/SAJ0876001I}, \href
  {https://ui.adsabs.harvard.edu/abs/2008SerAJ.176....1I} {176, 1}

\bibitem[\protect\citeauthoryear{{Jin}, {Gu}, {Huang}, {Shi}  \& {Feng}}{{Jin}
  et~al.}{2014}]{Jin2014}
{Jin} S.-W.,  {Gu} Q.,  {Huang} S.,  {Shi} Y.,   {Feng} L.-L.,  2014, \mn@doi
  [\apj] {10.1088/0004-637X/787/1/63}, \href
  {https://ui.adsabs.harvard.edu/abs/2014ApJ...787...63J} {787, 63}

\bibitem[\protect\citeauthoryear{{Kauffmann}, {White}, {Heckman}, {M{\'e}nard},
  {Brinchmann}, {Charlot}, {Tremonti}  \& {Brinkmann}}{{Kauffmann}
  et~al.}{2004}]{Kauffmann2004}
{Kauffmann} G.,  {White} S. D.~M.,  {Heckman} T.~M.,  {M{\'e}nard} B.,
  {Brinchmann} J.,  {Charlot} S.,  {Tremonti} C.,   {Brinkmann} J.,  2004,
  \mn@doi [\mnras] {10.1111/j.1365-2966.2004.08117.x}, \href
  {https://ui.adsabs.harvard.edu/abs/2004MNRAS.353..713K} {353, 713}

\bibitem[\protect\citeauthoryear{{Labb{\'e}} et~al.,}{{Labb{\'e}}
  et~al.}{2007}]{Labbe2007}
{Labb{\'e}} I.,  et~al., 2007, \mn@doi [\apj] {10.1086/519436}, \href
  {https://ui.adsabs.harvard.edu/abs/2007ApJ...665..944L} {665, 944}

\bibitem[\protect\citeauthoryear{{Laigle} et~al.,}{{Laigle}
  et~al.}{2016}]{Laigle2016}
{Laigle} C.,  et~al., 2016, \mn@doi [\apjs] {10.3847/0067-0049/224/2/24}, \href
  {http://adsabs.harvard.edu/abs/2016ApJS..224...24L} {224, 24}

\bibitem[\protect\citeauthoryear{{Laureijs}, {Duvet}, {Escudero Sanz},
  {Gondoin}, {Lumb}, {Oosterbroek}  \& {Saavedra Criado}}{{Laureijs}
  et~al.}{2010}]{Laureijs2010}
{Laureijs} R.~J.,  {Duvet} L.,  {Escudero Sanz} I.,  {Gondoin} P.,  {Lumb}
  D.~H.,  {Oosterbroek} T.,   {Saavedra Criado} G.,  2010, in Space Telescopes
  and Instrumentation 2010: Optical, Infrared, and Millimeter Wave. p. 77311H,
  \mn@doi{10.1117/12.857123}

\bibitem[\protect\citeauthoryear{{Lin} et~al.,}{{Lin} et~al.}{2014}]{Lin2014}
{Lin} L.,  et~al., 2014, \mn@doi [\apj] {10.1088/0004-637X/782/1/33}, \href
  {https://ui.adsabs.harvard.edu/abs/2014ApJ...782...33L} {782, 33}

\bibitem[\protect\citeauthoryear{{Lin}, {Fang}, {Cai}, {Wang}, {Fan}  \&
  {Kong}}{{Lin} et~al.}{2019}]{Lin2019}
{Lin} X.,  {Fang} G.,  {Cai} Z.-Y.,  {Wang} T.,  {Fan} L.,   {Kong} X.,  2019,
  \mn@doi [\apj] {10.3847/1538-4357/ab0e73}, \href
  {https://ui.adsabs.harvard.edu/abs/2019ApJ...875...83L} {875, 83}

\bibitem[\protect\citeauthoryear{{Ma{\'{\i}}z Apell{\'a}niz}}{{Ma{\'{\i}}z
  Apell{\'a}niz}}{2006}]{MaizApellaniz2006}
{Ma{\'{\i}}z Apell{\'a}niz} J.,  2006, \mn@doi [\aj] {10.1086/499158}, \href
  {http://adsabs.harvard.edu/abs/2006AJ....131.1184M} {131, 1184}

\bibitem[\protect\citeauthoryear{M\'{a}rmol-Queralt\'{o}, McLure, Cullen,
  Dunlop, Fontana  \& McLeod}{M\'{a}rmol-Queralt\'{o}
  et~al.}{2016}]{Marmol-Queralto2016}
M\'{a}rmol-Queralt\'{o} E.,  McLure R.~J.,  Cullen F.,  Dunlop J.~S.,  Fontana
  A.,   McLeod D.~J.,  2016, \mn@doi [Monthly Notices of the Royal Astronomical
  Society] {10.1093/mnras/stw1212}, 460, 3587

\bibitem[\protect\citeauthoryear{{McGee}, {Balogh}, {Wilman}, {Bower},
  {Mulchaey}, {Parker}  \& {Oemler}}{{McGee} et~al.}{2011}]{McGee2011}
{McGee} S.~L.,  {Balogh} M.~L.,  {Wilman} D.~J.,  {Bower} R.~G.,  {Mulchaey}
  J.~S.,  {Parker} L.~C.,   {Oemler} A.,  2011, \mn@doi [\mnras]
  {10.1111/j.1365-2966.2010.18189.x}, \href
  {https://ui.adsabs.harvard.edu/abs/2011MNRAS.413..996M} {413, 996}

\bibitem[\protect\citeauthoryear{{Mendel} et~al.,}{{Mendel}
  et~al.}{2015}]{Mendel2015}
{Mendel} J.~T.,  et~al., 2015, \mn@doi [\apjl] {10.1088/2041-8205/804/1/L4},
  \href {https://ui.adsabs.harvard.edu/abs/2015ApJ...804L...4M} {804, L4}

\bibitem[\protect\citeauthoryear{{Moresco} et~al.,}{{Moresco}
  et~al.}{2013}]{Moresco2013}
{Moresco} M.,  et~al., 2013, \mn@doi [\aap] {10.1051/0004-6361/201321797},
  \href {http://adsabs.harvard.edu/abs/2013A%26A...558A..61M} {558, A61}

\bibitem[\protect\citeauthoryear{{Oke} \& {Gunn}}{{Oke} \&
  {Gunn}}{1983}]{Oke1983}
{Oke} J.~B.,  {Gunn} J.~E.,  1983, \mn@doi [\apj] {10.1086/160817}, \href
  {http://adsabs.harvard.edu/abs/1983ApJ...266..713O} {266, 713}

\bibitem[\protect\citeauthoryear{{Peng} et~al.,}{{Peng}
  et~al.}{2010}]{Peng2010}
{Peng} Y.-j.,  et~al., 2010, \mn@doi [\apj] {10.1088/0004-637X/721/1/193},
  \href {http://adsabs.harvard.edu/abs/2010ApJ...721..193P} {721, 193}

\bibitem[\protect\citeauthoryear{{Polletta} et~al.,}{{Polletta}
  et~al.}{2007}]{Polletta2007}
{Polletta} M.,  et~al., 2007, \mn@doi [\apj] {10.1086/518113}, \href
  {https://ui.adsabs.harvard.edu/abs/2007ApJ...663...81P} {663, 81}

\bibitem[\protect\citeauthoryear{{Potter}, {Stadel}  \& {Teyssier}}{{Potter}
  et~al.}{2017}]{Potter2017}
{Potter} D.,  {Stadel} J.,   {Teyssier} R.,  2017, \mn@doi [Computational
  Astrophysics and Cosmology] {10.1186/s40668-017-0021-1}, \href
  {https://ui.adsabs.harvard.edu/abs/2017ComAC...4....2P} {4, 2}

\bibitem[\protect\citeauthoryear{{Pozzetti} \& {Mannucci}}{{Pozzetti} \&
  {Mannucci}}{2000}]{Pozzetti2000}
{Pozzetti} L.,  {Mannucci} F.,  2000, \mn@doi [\mnras]
  {10.1046/j.1365-8711.2000.03829.x}, \href
  {https://ui.adsabs.harvard.edu/abs/2000MNRAS.317L..17P} {317, L17}

\bibitem[\protect\citeauthoryear{{Renzini} \& {Peng}}{{Renzini} \&
  {Peng}}{2015}]{Renzini2015}
{Renzini} A.,  {Peng} Y.-j.,  2015, \mn@doi [\apj]
  {10.1088/2041-8205/801/2/L29}, \href
  {https://ui.adsabs.harvard.edu/abs/2015ApJ...801L..29R} {801, L29}

\bibitem[\protect\citeauthoryear{{Sanders} et~al.,}{{Sanders}
  et~al.}{2007}]{Sanders2007}
{Sanders} D.~B.,  et~al., 2007, \mn@doi [\apjs] {10.1086/517885}, \href
  {http://adsabs.harvard.edu/abs/2007ApJS..172...86S} {172, 86}

\bibitem[\protect\citeauthoryear{{Schweitzer} et~al.,}{{Schweitzer}
  et~al.}{2010}]{Schweitzer2010}
{Schweitzer} M.,  et~al., 2010, in Space Telescopes and Instrumentation 2010:
  Optical, Infrared, and Millimeter Wave. p. 77311K, \mn@doi{10.1117/12.857031}

\bibitem[\protect\citeauthoryear{{Scoville} et~al.,}{{Scoville}
  et~al.}{2007}]{Scoville2007}
{Scoville} N.,  et~al., 2007, \mn@doi [\apjs] {10.1086/516585}, \href
  {http://adsabs.harvard.edu/abs/2007ApJS..172....1S} {172, 1}

\bibitem[\protect\citeauthoryear{{Skrutskie} et~al.,}{{Skrutskie}
  et~al.}{2006}]{Skrutskie2006}
{Skrutskie} M.~F.,  et~al., 2006, \mn@doi [\aj] {10.1086/498708}, \href
  {http://adsabs.harvard.edu/abs/2006AJ....131.1163S} {131, 1163}

\bibitem[\protect\citeauthoryear{{Strateva} et~al.,}{{Strateva}
  et~al.}{2001}]{Strateva2001}
{Strateva} I.,  et~al., 2001, \mn@doi [\aj] {10.1086/323301}, \href
  {http://adsabs.harvard.edu/abs/2001AJ....122.1861S} {122, 1861}

\bibitem[\protect\citeauthoryear{{Wetzel}, {Tinker}  \& {Conroy}}{{Wetzel}
  et~al.}{2012}]{Wetzel2012}
{Wetzel} A.~R.,  {Tinker} J.~L.,   {Conroy} C.,  2012, \mn@doi [\mnras]
  {10.1111/j.1365-2966.2012.21188.x}, \href
  {https://ui.adsabs.harvard.edu/abs/2012MNRAS.424..232W} {424, 232}

\bibitem[\protect\citeauthoryear{{Wetzel}, {Tinker}, {Conroy}  \& {van den
  Bosch}}{{Wetzel} et~al.}{2013}]{Wetzel2013}
{Wetzel} A.~R.,  {Tinker} J.~L.,  {Conroy} C.,   {van den Bosch} F.~C.,  2013,
  \mn@doi [\mnras] {10.1093/mnras/stt469}, \href
  {https://ui.adsabs.harvard.edu/abs/2013MNRAS.432..336W} {432, 336}

\bibitem[\protect\citeauthoryear{{Whitaker} et~al.,}{{Whitaker}
  et~al.}{2011}]{Whitaker2011}
{Whitaker} K.~E.,  et~al., 2011, \mn@doi [\apj] {10.1088/0004-637X/735/2/86},
  \href {http://adsabs.harvard.edu/abs/2011ApJ...735...86W} {735, 86}

\bibitem[\protect\citeauthoryear{{Williams}, {Quadri}, {Franx}, {van Dokkum}
  \& {Labb{\'e}}}{{Williams} et~al.}{2009}]{Williams2009}
{Williams} R.~J.,  {Quadri} R.~F.,  {Franx} M.,  {van Dokkum} P.,   {Labb{\'e}}
  I.,  2009, \mn@doi [\apj] {10.1088/0004-637X/691/2/1879}, \href
  {http://adsabs.harvard.edu/abs/2009ApJ...691.1879W} {691, 1879}

\bibitem[\protect\citeauthoryear{{Wuyts} et~al.,}{{Wuyts}
  et~al.}{2007}]{Wuyts2007}
{Wuyts} S.,  et~al., 2007, \mn@doi [\apj] {10.1086/509708}, \href
  {http://adsabs.harvard.edu/abs/2007ApJ...655...51W} {655, 51}

\bibitem[\protect\citeauthoryear{{Wyder} et~al.,}{{Wyder}
  et~al.}{2007}]{Wyder2007}
{Wyder} T.~K.,  et~al., 2007, \mn@doi [\apjs] {10.1086/521402}, \href
  {http://adsabs.harvard.edu/abs/2007ApJS..173..293W} {173, 293}

\bibitem[\protect\citeauthoryear{{Zamojski} et~al.,}{{Zamojski}
  et~al.}{2007}]{Zamojski2007}
{Zamojski} M.~A.,  et~al., 2007, \mn@doi [\apjs] {10.1086/516593}, \href
  {http://adsabs.harvard.edu/abs/2007ApJS..172..468Z} {172, 468}

\bibitem[\protect\citeauthoryear{{Zehavi} et~al.,}{{Zehavi}
  et~al.}{2011}]{Zehavi2011}
{Zehavi} I.,  et~al., 2011, \mn@doi [\apj] {10.1088/0004-637X/736/1/59}, \href
  {http://adsabs.harvard.edu/abs/2011ApJ...736...59Z} {736, 59}

\makeatother
\end{thebibliography}
	
\vspace{10pt}
\vspace{10pt}
\noindent
$^{1}$ University of Nottingham, University Park, Nottingham NG7 2RD, UK\\
$^{2}$ INAF-Osservatorio di Astrofisica e Scienza dello Spazio di Bologna, Via Piero Gobetti 93/3, I-40129 Bologna, Italy\\
$^{3}$ INAF-Osservatorio Astronomico di Brera, Via Brera 28, I-20122 Milano, Italy\\
$^{4}$ Observatoire Astronomique de Strasbourg (ObAS), Universit\'e de Strasbourg - CNRS, UMR 7550, Strasbourg, France\\
$^{5}$ INAF-IASF Milano, Via Alfonso Corti 12, I-20133 Milano, Italy\\
$^{6}$ Instituto de Astrof\'isica e Ci\^encias do Espa\c{c}o, Universidade do Porto, CAUP, Rua das Estrelas, PT4150-762 Porto, Portugal\\
$^{7}$ Institute of Cosmology and Gravitation, University of Portsmouth, Portsmouth PO1 3FX, UK\\
$^{8}$ Dipartimento di Fisica e Astronomia, Universit\'a di Bologna, Via Gobetti 93/2, I-40129 Bologna, Italy\\
$^{9}$ INAF-Osservatorio Astrofisico di Arcetri, Largo E. Fermi 5, I-50125, Firenze, Italy\\
$^{10}$ INAF-Osservatorio Astrofisico di Torino, Via Osservatorio 20, I-10025 Pino Torinese (TO), Italy\\
$^{11}$ Institut de F\'isica d'Altes Energies IFAE, 08193 Bellaterra, Barcelona, Spain\\
$^{12}$ Institute of Space Sciences (ICE, CSIC), Campus UAB, Carrer de Can Magrans, s/n, 08193 Barcelona, Spain\\
$^{13}$ Institut d'Estudis Espacials de Catalunya (IEEC), 08034 Barcelona, Spain\\
$^{14}$ INAF-Osservatorio Astronomico di Roma, Via Frascati 33, I-00078 Monteporzio Catone, Italy\\
$^{15}$ Department of Physics "E. Pancini", University Federico II, Via Cinthia 6, I-80126, Napoli, Italy\\
$^{16}$ INFN section of Naples, Via Cinthia 6, I-80126, Napoli, Italy\\
$^{17}$ INAF-Osservatorio Astronomico di Capodimonte, Via Moiariello 16, I-80131 Napoli, Italy\\
$^{18}$ Centre National d'Etudes Spatiales, Toulouse, France\\
$^{19}$ Institute for Astronomy, University of Edinburgh, Royal Observatory, Blackford Hill, Edinburgh EH9 3HJ, UK\\
$^{20}$ ESAC/ESA, Camino Bajo del Castillo, s/n., Urb. Villafranca del Castillo, 28692 Villanueva de la Ca\~nada, Madrid, Spain\\
$^{21}$ Mullard Space Science Laboratory, University College London, Holmbury St Mary, Dorking, Surrey RH5 6NT, UK\\
$^{22}$ INFN-Padova, Via Marzolo 8, I-35131 Padova, Italy\\
$^{23}$ INAF-Osservatorio Astronomico di Trieste, Via G. B. Tiepolo 11, I-34131 Trieste, Italy\\
$^{24}$ von Hoerner \& Sulger GmbH, Schlo{\ss}Platz 8, D-68723 Schwetzingen, Germany\\
$^{25}$ Universit\"ats-Sternwarte M\"unchen, Fakult\"at f\"ur Physik, Ludwig-Maximilians-Universit\"at M\"unchen, Scheinerstrasse 1, 81679 M\"unchen, Germany\\
$^{26}$ Max-Planck-Institut f\"ur Astronomie, K\"onigstuhl 17, D-69117 Heidelberg, Germany\\
$^{27}$ Aix-Marseille Univ, CNRS/IN2P3, CPPM, Marseille, France\\
$^{28}$ Institut de Physique Nucl\'eaire de Lyon, 4, rue Enrico Fermi, 69622, Villeurbanne cedex, France\\
$^{29}$ Universit\'e de Gen\`eve, D\'epartement de Physique Th\'eorique and Centre for Astroparticle Physics, 24 quai Ernest-Ansermet, CH-1211 Gen\`eve 4, Switzerland\\
$^{30}$ Aix-Marseille Univ, CNRS, CNES, LAM, Marseille, France\\
$^{31}$ Institute of Theoretical Astrophysics, University of Oslo, P.O. Box 1029 Blindern, N-0315 Oslo, Norway\\
$^{32}$ Argelander-Institut f\"ur Astronomie, Universit\"at Bonn, Auf dem H\"ugel 71, 53121 Bonn, Germany\\
$^{33}$ Centre for Extragalactic Astronomy, Department of Physics, Durham University, South Road, Durham, DH1 3LE, UK\\
$^{34}$ Jet Propulsion Laboratory, California Institute of Technology, 4800 Oak Grove Drive, Pasadena, CA, 91109, USA\\
$^{35}$ University of Paris Denis Diderot, University of Paris Sorbonne Cit\'e (PSC), 75205 Paris Cedex 13, France\\
$^{36}$ Sorbonne Universit\'e, Observatoire de Paris, Universit\'e PSL, CNRS, LERMA, F-75014, Paris, France\\
$^{37}$ Institut d'Astrophysique de Paris, 98bis Boulevard Arago, F-75014, Paris, France\\
$^{38}$ IRFU, CEA, Universit\'e Paris-Saclay F-91191 Gif-sur-Yvette Cedex, France\\
$^{39}$ Observatoire de Sauverny, Ecole Polytechnique F\'ed\'erale de Lau- sanne, CH-1290 Versoix, Switzerland\\
$^{40}$ Department of Astronomy, University of Geneva, ch. d'\'Ecogia 16, CH-1290 Versoix, Switzerland\\
$^{41}$ AIM, CEA, CNRS, Universit\'{e} Paris-Saclay, Universit\'{e} Paris Diderot, Sorbonne Paris Cit\'{e}, F-91191 Gif-sur-Yvette, France\\
$^{42}$ Space Science Data Center, Italian Space Agency, via del Politecnico snc, 00133 Roma, Italy\\
$^{43}$ Max Planck Institute for Extraterrestrial Physics, Giessenbachstr. 1, D-85748 Garching, Germany\\
$^{44}$ Institute of Space Sciences (IEEC-CSIC), c/Can Magrans s/n, 08193 Cerdanyola del Vall\'es, Barcelona, Spain\\
$^{45}$ Instituto de Astrof\'isica e Ci\^encias do Espa\c{c}o, Faculdade de Ci\^encias, Universidade de Lisboa, Tapada da Ajuda, PT-1349-018 Lisboa, Portugal\\
$^{46}$ Departamento de F\'isica, Faculdade de Ci\^encias, Universidade de Lisboa, Edif\'icio C8, Campo Grande, PT1749-016 Lisboa, Portugal\\
$^{47}$ Universidad Polit\'ecnica de Cartagena, Departamento de Electr\'onica y Tecnolog\'ia de Computadoras, 30202 Cartagena, Spain\\
$^{48}$ Istituto Nazionale di Fisica Nucleare, Sezione di Bologna, Via Irnerio 46, I-40126 Bologna, Italy\\
$^{49}$ Infrared Processing and Analysis Center, California Institute of Technology, Pasadena, CA 91125, USA\\

	\appendix

	\section{Comparison among data sets}\label{sec:mag_data sets}
	In this appendix, we compare the relevant properties of galaxies in the different data sets considered in this work. Results are shown at \Euclid Wide Survey depth. \par
\begin{figure*}
		\centering
		\includegraphics[width=1\linewidth, keepaspectratio]{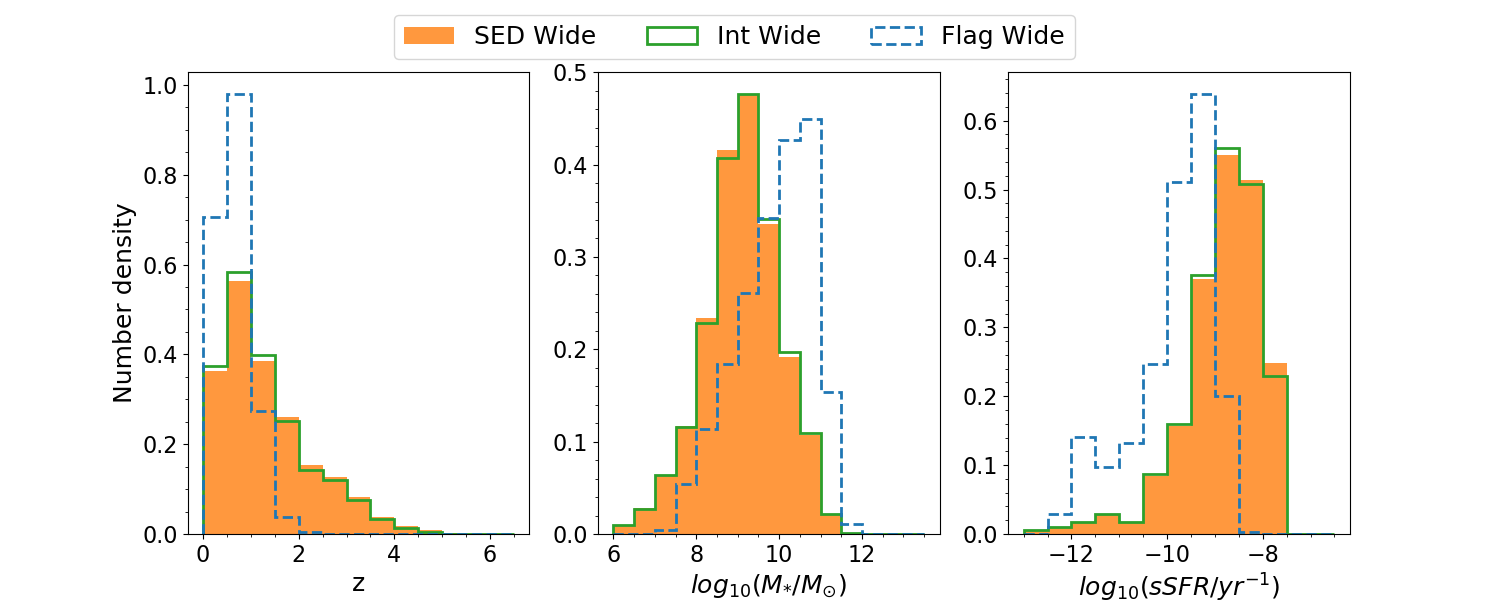}
		\caption{Distribution of redshift (\textit{left}), stellar mass (\textit{centre}), and sSFR (\textit{right}) for galaxies in the three different data sets considered in this work: \textit{SED Wide} (\textit{filled orange histograms}), \textit{Int Wide} (\textit{green solid lines}), and \textit{Flag Wide} (\textit{blue dashed lines}).}
		\label{fig:data set_galprop}
	\end{figure*}

	\autoref{fig:data set_galprop} shows the redshift, stellar mass, and sSFR distribution of the \textit{SED Wide}, \textit{Int Wide}, and \textit{Flag Wide} data sets. The first two data sets show similar galaxy properties, as expected given that they are derived from the same parent sample of real galaxies. This confirms that the different model and photometric error assumptions are not affecting the results. The \textit{Flag Wide} data set is limited to galaxies at $z\lesssim2$ with generally larger stellar mass and lower star-formation than the other two data sets. We verify that the difference in the stellar mass and sSFR distributions are not entirely caused by the difference in the redshift distributions and are indeed still present even in low-redshift galaxies. \par
	
\begin{figure*}
		\centering
		\includegraphics[width=0.7\linewidth, keepaspectratio]{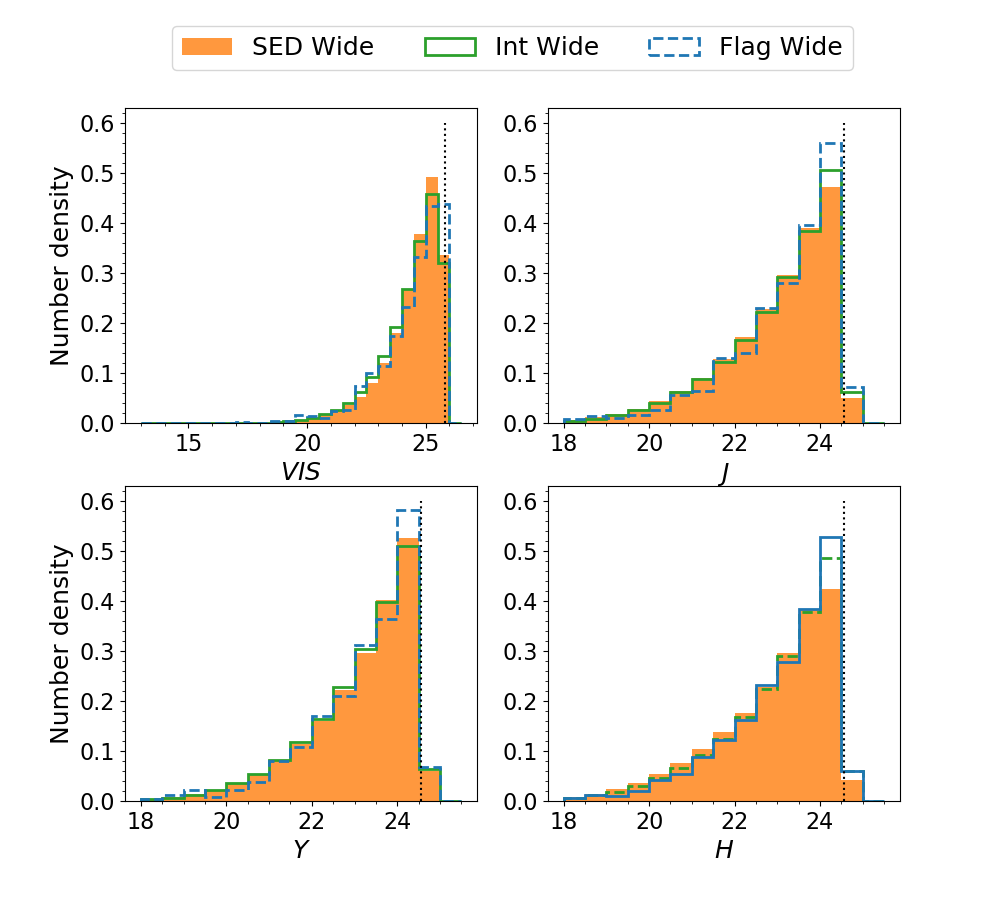}
		\caption{Distribution of magnitudes in the $VIS$ (\textit{top left}), $J$ (\textit{top right}), $Y$ (\textit{bottom left}), and $H$ (\textit{bottom right}) bands for galaxies in the three different data sets considered in this work: \textit{SED Wide} (\textit{filled orange histograms}), \textit{Int Wide} (\textit{green solid lines}), and \textit{Flag Wide} (\textit{blue dashed lines}).The vertical dotted lines indicate the magnitude corresponding to a S/N$=$3 for each filter. }
		\label{fig:data set_mag}
	\end{figure*}

	\autoref{fig:data set_mag} shows the magnitude distribution of galaxies in the \Euclid filters for the three data sets with the depth of the \Euclid Wide Survey. The two data sets derived from real galaxies, i.e., \textit{SED Wide} and \textit{Int Wide}, have similar magnitude distributions in the \Euclid filters. Mock galaxies in the \textit{Flag Wide} data set have instead fainter $VIS$ band magnitudes, as a possible consequence of galaxies being less star-forming in this data set. The magnitudes in the other \Euclid filters are instead similar among the three different data sets.
	
	\bsp	
	\label{lastpage}
\end{document}